\documentclass[12pt,draftclsnofoot,onecolumn]{IEEEtran}

\makeatletter
\def\bstctlcite{\@ifnextchar[{\@bstctlcite}{\@bstctlcite[@auxout]}}
\def\@bstctlcite[#1]#2{%
 \@bsphack
 \@for\@citeb:=#2\do{%
 \edef\@citeb{\expandafter\@firstofone\@citeb}%
 \if@filesw\immediate\write\csname #1\endcsname{\string\citation{\@citeb}}\fi}%
 \@esphack}
\makeatother
\IEEEoverridecommandlockouts

\usepackage[T1]{fontenc}
\usepackage[english]{babel}
\usepackage{cite}
\ifCLASSINFOpdf
  \usepackage[pdftex]{graphicx}
\else
\fi
\usepackage{amsmath}
\usepackage[cmintegrals]{newtxmath}
\usepackage{algorithm}
\usepackage{algorithmic}
\usepackage{array}
\usepackage{float}
\usepackage[caption=false,font=footnotesize]{subfig}
\usepackage{stfloats}
\usepackage{url}
\usepackage{dane_nmath}
\usepackage{tikz}
\usepackage{ellipsis}
\usetikzlibrary{calc}
\usetikzlibrary{decorations.pathreplacing,decorations.markings,shapes.geometric}
\usetikzlibrary{chains}

\tikzstyle{block} = [draw, rectangle, 
    minimum height=4em, minimum width=6em]
\tikzstyle{input} = [coordinate]
\tikzstyle{output} = [coordinate]
\tikzstyle{pinstyle} = [pin edge={to-,thin,black}]

\usetikzlibrary{positioning}

\tikzset{radiation/.style={{decorate,decoration={expanding waves,angle=90,segment length=5pt}}}}
\tikzset{font={\fontsize{10pt}{12}\selectfont}}

\usetikzlibrary{spy}
\usepackage{pgfplots}
\usepackage{wrapfig}
\usetikzlibrary{arrows,shapes}
\usetikzlibrary{positioning,shapes.callouts}
\usepgfplotslibrary{groupplots,dateplot}
\usetikzlibrary{patterns,shapes.arrows}
\pgfplotsset{compat=newest}
\def\BibTeX{{\rm B\kern-.05em{\sc i\kern-.025em b}\kern-.08em
    T\kern-.1667em\lower.7ex\hbox{E}\kern-.125emX}}

\newcommand{\vari}[1]{\textcolor{black}{#1}}

\newcommand\copyrighttext{%
	\footnotesize \textcopyright This work has been submitted to the IEEE for possible publication. Copyright may be transferred without notice, after which this version may no longer be accessible.}
\newcommand\copyrightnotice{%
	\begin{tikzpicture}[remember picture,overlay]
	\node[anchor=south,yshift=10pt] at (current page.south) {\fbox{\parbox{\dimexpr\textwidth-\fboxsep-\fboxrule\relax}{\copyrighttext}}};
	\end{tikzpicture}%
}

\begin{document}
\bstctlcite{IEEEexample:BSTcontrol}
\title{Learning the CSI Recovery in FDD Systems}

%
%
%

\author{Wolfgang~Utschick,~\IEEEmembership{Fellow},~IEEE, Valentina~Rizzello,~\IEEEmembership{Student Member,~IEEE,}
        Michael~Joham,~\IEEEmembership{Member,~IEEE,}
        \\
        Zhengxiang~Ma,~\IEEEmembership{Member,~IEEE,}
        and~Leonard~Piazzi,~\IEEEmembership{Member,~IEEE} \\ 
\thanks{W.~Utschick, V.~Rizzello and M.~Joham are with the Professur f{\"u}r
Methoden der Signalverarbeitung, Technische Universit{\"a}t M{\"u}nchen, Munich,
80333, Germany. \{\texttt{utschick, valentina.rizzello, joham}\}\texttt{@tum.de}}
\thanks{Z.~Ma and L.~Piazzi are with Futurewei Technologies, at 400 Crossing Blvd., Bridgewater, New Jersey 08807, USA. \{\texttt{zma@futurewei.com, lpiazzi@verizon.net}\}}
}

\maketitle

\copyrightnotice

\begin{abstract}
  We propose an innovative machine learning-based technique to address the
  problem of channel acquisition at the base station in frequency division duplex systems. In this context, the base station reconstructs the full channel state information in the downlink frequency range based on limited downlink channel state information feedback from the mobile terminal. The channel state information recovery is based on a convolutional neural network which is trained exclusively on collected channel state samples acquired in the uplink frequency domain. 
No acquisition of training samples in the downlink frequency range is required at all. Finally, after a detailed presentation and analysis of the proposed technique and its performance, the ``transfer learning'' assumption of the convolutional neural network that is central to the proposed approach is validated with an analysis based on the maximum mean discrepancy metric.
\end{abstract}

\begin{IEEEkeywords}
Machine learning, massive MIMO, FDD systems, transfer learning, maximum mean discrepancy, convolutional neural networks, deep learning.
\end{IEEEkeywords}

%
\IEEEpeerreviewmaketitle

\section{Introduction}
\label{sec: intro}
\IEEEPARstart{T}{he} massive multiple-input multiple-output (MIMO) technology is one of the most prominent directions to scale up capacity and throughput in modern communication systems \cite{marzetta}. In particular, the
multi-antennas support at the base station (BS) makes simple techniques such as
spatial multiplexing and beamforming very efficient regarding the spectrum or
the bandwidth utilization. However, to take full advantage of Massive MIMO systems, the base station must have the best possible channel estimation. Considering the typically stringent delay requirements in wireless mobile communication systems, the channel state information (CSI) has to be acquired in very short regular time intervals. \\
A variety of solution approaches developed for this purpose are based on time division duplex (TDD) mode. TDD means that both the BS and the mobile terminal (MT) share the same bandwidth, but the BS and the MT cannot transmit in the same time
interval. Due to the fact that they both share the same bandwidth, the uplink (UL) and downlink (DL) channels are reciprocal, i.e., once the BS estimates the UL channel, it also knows the DL channel without any additional feedback overhead \cite{sanguinetti2019massive}. \\ On the other hand, in frequency
division duplex (FDD) mode, the BS and the MT transmit in the same time slot but
at different frequencies. This breaks the reciprocity between UL CSI and DL CSI and
makes it hard for the network operators with FDD licenses to obtain an accurate
DL CSI estimate for transmit signal processing \cite{2019massive}. The usual solution to the problem is to either extrapolate the DL CSI from the estimate of the UL CSI at the BS, or to transfer the DL CSI estimated at the MT to the BS directly or in a highly compressed version.\footnote{Another common solution is to omit the feedback of the CSI and to signal only channel quality indicators of the transmission properties.} The former rely on available system models and aim to rather estimate the second-order information of the DL channel, namely the covariance matrix of the DL CSI, which is supposed to stay constant for several coherence intervals. See \cite{8542957} for latest results in this direction and \cite{UtNo99} for a reference to a very early attempt by the author. In contrast, data-driven approaches do not make assumptions about the underlying channel model, but instead use paired training samples of UL CSI and DL CSI along with a machine learning procedure that predicts the DL CSI considering the UL CSI as input \cite{ArDoCaYaHoBr19,alk2019deep,8764345,han2020deep,safari,me}.
In the end, the most common solutions encountered in practice are based on feedback-based methods. In addition to a plethora of classical approaches, cf. \cite{4641946}, a promising variant that has recently been proposed several times in different versions is to elegantly combine the learning of a sparse feedback format with the task of performing channel reconstruction at the BS using an autoencoder neural network \cite{8322184,8638509,9090892,8972904,9279228,9347820}. To this end, these approaches rely on implementing the encoder and decoder part of the jointly trained autoencoder distributed on MT and the BS, i.e., whereas the encoder unit at the MT maps channel estimates or corresponding observations of pilot data to an appropriate feedback format, the decoder unit at the BS reconstructs the complete DL channel estimates based on the received feedback.
\vari{The approach in~\cite{Liang2020} instead proposes a centralized training at the BS station based on DL CSI data, where the DL CSI is compressed at the user side using a Gaussian random matrix and it is then fed back at the BS which reconstructs it with a neural network.
In~\cite{Mashhadi2021b} a model-based neural network has been trained to jointly design the pilot pattern and estimate the DL CSI in FDD MIMO orthogonal frequency division multiplex systems (OFDM).}\\
\vari{The approach proposed in this paper also belongs to the data-driven category. However, compared to other approaches found in the literature, we consider that in {real-world systems} (i) the implementation of a distributed training setup between BS and MT is rather challenging in practice, (ii) a centralized UL-DL CSI based supervised training would require a large number of true UL-DL CSI \emph{pairs}, which are very costly to obtain at the same location - usually the BS, (iii) and thus a centralized training at the BS using true DL CSI data would impose an unrealistic overhead, as transporting the data is particularly inefficient since it leads to excessive network usage.}
\vari{With our contribution:
\begin{enumerate}
    \item the practical issue of ``DL CSI data acquisition'' by training a neural network at the BS in a ``centralized fashion'' using only UL CSI data available at the BS is addressed;
    \item the typical overhead of the "federated" learning framework, where the MT would have to send the gradients to the BS several times during training to update the neural network parameters, is avoided;
    \item a new perspective to FDD systems is presented, where we show that the BS can obtain the DL CSI estimate with a learning which is solely based on UL CSI data;
    \item a justification of the proposed ``UL-DL conjecture'' using the maximum-mean-discrepancy metric is given;
    \item and a neural network approach which consists on convolutional layers to show that the proposed ``UL-DL conjecture'' works is presented. Please keep in mind that the neural network architecture we used, which we kept as simple as possible, is only a means to show that our idea works in general. Despite the already promising performance, further optimizations of the architecture are the subject of future research.
\end{enumerate}}

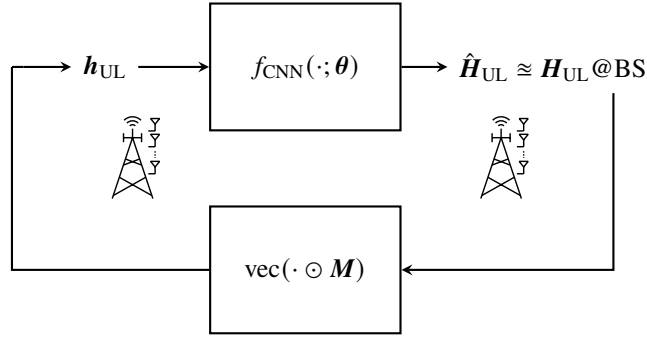
\begin{figure}[!t]
    \centering
    \definecolor{mycolor}{HTML}{7B241C}

\def\mobilestation{
\begin{tikzpicture}[line width=1.1pt]
 \draw
 (1.3, -1.3) -- (1.3, -0.3)
 (1.9, -1.3) -- (1.9, -0.3)
 (1.9, -0.3) -- (1.3, -0.3)
 (1.3, -1.3) -- (1.9, -1.3);
  \draw (1.3, -0.4) -- (1.0, -0.4) (1.0, -0.4)--(1.0, -0.1) (1.0, -0.1)--(0.8, 0.15) (1.0, -0.1)--(1.2, 0.15) (1.2, 0.15)--(0.8, 0.15);
\end{tikzpicture}
}

\def\basestation{
\begin{tikzpicture}[line width=1.8pt]
 \draw
    (-0.4, 0.2) -- (-0.4, -0.2)
    (0.4, 0.2) -- (0.4, -0.2)
    (0, 0.4) -- (0, 0)
    (-0.4, 0) -- (0.4, 0)
    (-0.9, -2.7) -- (0, 0) coordinate[pos=0.05] (l3)
    coordinate[pos=0.35] (l4)
        coordinate[pos=0.55] (l2) coordinate[pos=0.65] (l1)
    (0.9, -2.7) -- (0, 0) coordinate[pos=0.05] (r3)
    coordinate[pos=0.35] (r4)
        coordinate[pos=0.55] (r2) coordinate[pos=0.65] (r1);
  \draw (l1) -- (r2) (l2)--(r1) (l3)--(r4) (r3)--(l4);
  \draw[radiation,decoration={angle=45}] (0, 0.4) -- ++ (90:0.7cm);
  \draw (0.7, -0.4) -- (1.0, -0.4) (1.0, -0.4)--(1.0, -0.1) (1.0, -0.1)--(0.8, 0.15) (1.0, -0.1)--(1.2, 0.15) (1.2, 0.15)--(0.8, 0.15);
  \draw (0.7, 0.3) -- (1.0, 0.3) (1.0, 0.3)--(1.0, 0.6) (1.0, 0.6)--(0.8, 0.85) (1.0, 0.6)--(1.2, 0.85) (1.2, 0.85)--(0.8, 0.85);
  \draw[dotted] (1.0, -0.6) -- (1.0, -0.9);
  \draw (0.7, -1.6) -- (1.0, -1.6) (1.0, -1.6)--(1.0, -1.3) (1.0, -1.3)--(0.8, -1.05) (1.0, -1.3)--(1.2, -1.05) (1.2, -1.05)--(0.8, -1.05);
\end{tikzpicture}
}

\tikzset{>=latex}

\begin{tikzpicture}[auto, node distance=2.7cm, line width=0.8pt]
    \node [input, name=input] {};
    \node [block, right of=input,
        node distance=3.9cm] (cnn) {$f_\text{CNN}(\cdot;{\boldsymbol{\theta}})$};
    \node [output, right of=cnn] (output) {};
    \node[below right=-0.3cm and 1cm of cnn, scale=0.3]{
    \basestation
    };
    \node [block, below of=cnn] (mask) {$\vectoriz(\cdot \odot \vM)$};
    \node[below right=-0.3cm and -3.9cm of cnn, scale=0.3]{
    \basestation
    };
    \node(n1){};
    \path (cnn) -- node [right=-0.1cm of cnn] (text2) {$\ulbighat \approxeq \ulbig$@BS}(output);
    \path (n1) -- node[right=-0.6cm of n1] (text1) {$\ulpartial$} (cnn);
    
    \draw [-] (mask) -| node[pos=0.99] {} node [near end] {} (input);
    \draw[->, >=stealth] (n1)--(text1);
    \draw [-] (n1) -| node[pos=0.99]{}node [near end] {} (input);
    \draw[->, >=stealth] (text1)--(cnn);
    \draw[->, >=stealth] (cnn)--(text2);
    \node(n3) [below right=-0.3cm and -0.75cm of text2]{};
    \draw [->, >=stealth] (n3) |- (mask);
\end{tikzpicture}
    \caption{CNN training based on paired UL CSI $(\ulpartial,\ulbig)$ collected at the BS.}
    \label{fig: ul_training_blk}
\end{figure}

\noindent In the following, we present the principle of the training based solely on UL CSI. By denoting $\ulbig$ and $ \dlbig
\in \mathbb{C}^{N_{\text{a}} \times N_{\text{c}} }$ as the true uplink  and downlink channel matrices, respectively,
where $N_{\text{a}}$ and $N_{\text{c}}$ denote the number of antennas at the BS and the
number of subcarriers, respectively, we can summarize our approach in two
stages. \\
Firstly, we train a convolutional neural network (CNN) at the BS to reconstruct the full UL channel matrix from a low-sampled version of itself, cf. Figure \ref{fig: ul_training_blk}.
This stage can be formulated as
\begin{equation}
  \ulbighat = f_\text{CNN}(\ulpartial;{\boldsymbol{\theta}}),
  \label{eq:ULCNN}
\end{equation}
where $\ulbighat \in \mathbb{C}^{N_{\text{a}} \times N_{\text{c}} }$ denotes
the reconstructed UL CSI, $f_\text{CNN}(\cdot;{\boldsymbol{\theta}})$ denotes the function instantiated by the CNN, and $\ulpartial$ represents the low-sampled version of the true UL channel matrix. Note that the training phase of the CNN at the BS is solely based on collected CSI estimates in the UL frequency range. \\ 
\begin{figure}[!t]
    \centering
    \definecolor{mycolor}{HTML}{7B241C}

\def\mobilestation{
\begin{tikzpicture}[line width=1.1pt]
 \draw
 (1.3, -1.3) -- (1.3, -0.3)
 (1.9, -1.3) -- (1.9, -0.3)
 (1.9, -0.3) -- (1.3, -0.3)
 (1.3, -1.3) -- (1.9, -1.3);
  \draw (1.3, -0.4) -- (1.0, -0.4) (1.0, -0.4)--(1.0, -0.1) (1.0, -0.1)--(0.8, 0.15) (1.0, -0.1)--(1.2, 0.15) (1.2, 0.15)--(0.8, 0.15);
\end{tikzpicture}
}

\def\basestation{
\begin{tikzpicture}[line width=1.8pt]
 \draw
    (-0.4, 0.2) -- (-0.4, -0.2)
    (0.4, 0.2) -- (0.4, -0.2)
    (0, 0.4) -- (0, 0)
    (-0.4, 0) -- (0.4, 0)
    (-0.9, -2.7) -- (0, 0) coordinate[pos=0.05] (l3)
    coordinate[pos=0.35] (l4)
        coordinate[pos=0.55] (l2) coordinate[pos=0.65] (l1)
    (0.9, -2.7) -- (0, 0) coordinate[pos=0.05] (r3)
    coordinate[pos=0.35] (r4)
        coordinate[pos=0.55] (r2) coordinate[pos=0.65] (r1);
  \draw (l1) -- (r2) (l2)--(r1) (l3)--(r4) (r3)--(l4);
  \draw[radiation,decoration={angle=45}] (0, 0.4) -- ++ (90:0.7cm);
  \draw (0.7, -0.4) -- (1.0, -0.4) (1.0, -0.4)--(1.0, -0.1) (1.0, -0.1)--(0.8, 0.15) (1.0, -0.1)--(1.2, 0.15) (1.2, 0.15)--(0.8, 0.15);
  \draw (0.7, 0.3) -- (1.0, 0.3) (1.0, 0.3)--(1.0, 0.6) (1.0, 0.6)--(0.8, 0.85) (1.0, 0.6)--(1.2, 0.85) (1.2, 0.85)--(0.8, 0.85);
  \draw[dotted] (1.0, -0.6) -- (1.0, -0.9);
  \draw (0.7, -1.6) -- (1.0, -1.6) (1.0, -1.6)--(1.0, -1.3) (1.0, -1.3)--(0.8, -1.05) (1.0, -1.3)--(1.2, -1.05) (1.2, -1.05)--(0.8, -1.05);
\end{tikzpicture}
}

\tikzset{>=latex}

\begin{tikzpicture}[auto, node distance=2.7cm, line width=0.8pt]
    \node [input, name=input] {};
    \node [block, right of=input,
        node distance=5cm] (cnn) {$f_\text{CNN}(\cdot;{\boldsymbol{\theta}})$};
    \node [output, right of=cnn] (output) {};
    \node[below right=-0.3cm and 1cm of cnn, scale=0.3]{
    \basestation
    };
    \node [block, below of=cnn] (mask) {$\vectoriz(\cdot \odot \vM)$};
    \node[below right=-0.4cm and 1cm of mask, scale=0.6]{
    \mobilestation
    };
    \node[below right=-0.3cm and -5cm of cnn, scale=0.3]{
    \basestation
    };
    \node[below right=2cm and -5.5cm of cnn, scale=1]{
    $ \substack{\displaystyle\text{partial CSI} \\ \displaystyle \text{feedback}} $};
    \node(n1){};
    \path (cnn) -- node [right=-0.1cm of cnn] (text2) {$\dlbighat$@BS}(output);
    \path (n1) -- node[right=-1cm of n1] (text1) {$\dlfeedb$@BS} (cnn);
    \draw [-] (mask) -| node[pos=0.99] {} node [near end] {} (input);
    \draw[->, >=stealth] (n1)--(text1);
    \draw [-] (n1) -| node[pos=0.99]{}node [near end] {} (input);
    \draw[->, >=stealth] (text1)--(cnn);
    \draw[->, >=stealth] (cnn)--(text2);
    \node(n3) [right=0.6cm of mask]{$\dlbig$@MT};
    \draw [->, >=stealth] (n3) -- (mask);
\end{tikzpicture}
    \caption{DL CSI recovery based on partial DL CSI from MT.}
    \label{fig: dl_prediction_blk}
\end{figure}
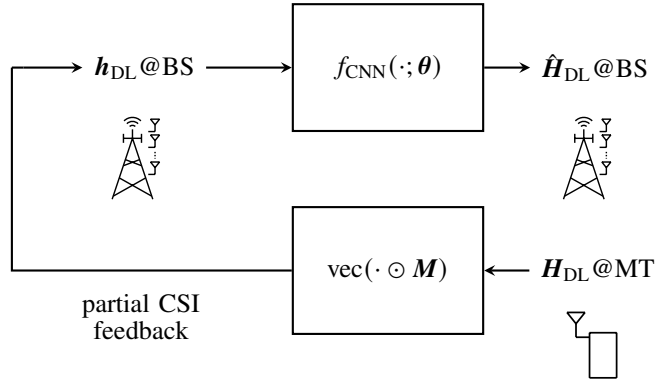
\noindent In the second stage, we assume that the MT, which has access to the full DL channel matrix,\footnote{Without any restriction on the principle, it would also be conceivable that the channel estimation in the MT is limited only to the part of the full DL channel matrix that is considered for feedback.} feeds the low-sampled version of it back to the BS, where this represents the required input of the trained CNN at the BS in order to reconstruct the full DL channel matrix, cf. Figure \ref{fig: dl_prediction_blk}. We can summarize this stage as
\begin{equation}
  \dlbighat = f_\text{CNN}(\dlfeedb;{\boldsymbol{\theta}}),
\end{equation}
where $\dlbighat \in \mathbb{C}^{N_{\text{a}} \times N_{\text{c}} }$ denotes
the reconstructed DL channel matrix, as in (\ref{eq:ULCNN}), $f_\text{CNN}(\cdot;{\boldsymbol{\theta}})$ still denotes the function instantiated by the UL-trained CNN, and
$\dlfeedb$ contains the low-sampled version of the true DL channel matrix with the same format and size as in the first stage. \\
The proposed technique is obviously based on the conjecture that learning the reconstruction in the UL domain can be ``transfered'' over the frequency gap between UL and DL center frequencies to the DL domain without any further adaptation.
\\
The rest of the paper is organized as follows. First, in Section~\ref{sec: scenario}, we describe how the channel dataset is constructed. Then, in Section~\ref{sec: learn_ul}, we present
the details of the UL training procedure, while in Section~\ref{sec: dl_pred}, we
deal with the reconstruction of the DL CSI. The obtained results are discussed in
Section~\ref{sec: result_single}. In Section~\ref{sec: cnn_on_another_cell}, we
evaluate how the learned CNN performs on another cell. Finally, in
Section~\ref{sec: verify_conj} with an analysis based on the maximum mean discrepancy (MMD) metric, we justify the stated conjecture and its results from a statistical point of view.

\section{Scenario and dataset description}
\label{sec: scenario}
\noindent The following study is based on an FDD system that utilizes center frequencies below
$6$~GHz and we explore three different frequency gaps between UL and DL, namely
$120$~MHz, $240$~MHz, and $480$~MHz. The channel state information for the UL and DL scenario has been generated with the $\Matlab$
based channel simulator QuaDRiGa version 2.2~\cite{quad, quad2}. \\
We simulate an urban microcell (UMi) non-line-of-sight (NLoS) scenario, where the
number of multi-path components (MPCs) is $L=58$.
The BS is placed at a height of $10$ meters and is equipped with a
uniform planar array (UPA) with $N_{\text{a}} =  8\times 8$ ``3GPP-3d'' antennas, while the
users have a single omni-directional antenna each. Additionally, the BS antennas
are tilted by $6$ degrees towards the ground to point in the direction of the
users. \\

\noindent We consider a bandwidth of approximately $8$~MHz divided over $N_{\text{c}} =  160$ subcarriers. The UL center frequency is $2.5$~GHz while the DL center frequencies are
$2.62$~GHz, $2.74$~GHz, and $2.98$~GHz. 
The radio propagation characteristic of the entire cell which supports a radius of
$150$~m has been uniformly sampled at $3\times 10^5$ different locations and for each sample, the channels at the
predefined frequency ranges are collected. Consequently, the dataset is split into three groups of $2,4\times 10^5$, $3\times 10^4$ and $3\times 10^4$ samples,
where each sample consists of the four matrices ${\vH}_{\text{UL}} $, ${\vH}_{\text{DL-120}} $, ${\vH}_{\text{DL-240}} $ and ${\vH}_{\text{DL-480}} \in \mathbb{C}^{N_{\text{a}}\times N_{\text{c}}} $.
Note that since our training is based on the UL CSI exclusively, only the test 
set of the three DL CSI datasets (\texttt{DL@120...480}) will be used.\\
In order to maintain the spatial consistency of the scenario, for a given
environment, the following parameters are identical in UL and DL domain: positions of BS and MTs, propagation delays and angles for each MPC, and large scale fading parameters. Small scale parameters are chosen independently, consistent with the stochastic nature of phase shifts associated with MPCs as frequency changes, due to which exptrapolations over the UL-DL frequency gap are hardly possible. Here, the proposed solution approach based on mere reconstruction in the same bandwidth proves to be particularly advantageous. \\
Within QuaDRiGa, the channel between the $N_{\text{a}}$ transmit antennas and the single receive antenna at the MS is modeled as
$$ [\vect{H}]_n = \sum_{\ell=1}^L \vect{g}_\ell \exp(-j 2\pi f_n \tau_\ell), $$
corresponding to the $n$-th column vector of the introduced channel matrix. The parameters are the $n$-th carrier frequency $f_n$ of $N_{\text{c}} $ carriers, the time delay $\tau_\ell$ of the $\ell$-th of a total of $ L$ paths and $\vect{g}_\ell$ as the channel vector consisting of the complex-valued channel gains $g_{k, \ell}$ of the $\ell$-th path between the $k$-th transmit antenna and the receive antenna at the MS, 
depending on the polarimetric antenna responses at the receiver and the transmitter and on the arrival and departure angles $(\phi_\ell^\text{a}, \theta_\ell^\text{a})$ and $(\phi_\ell^\text{d}, \theta_\ell^\text{d})$.\footnote{Azimuthal and elevation angles with respect to the array geometry under consideration.} 
\\
The dataset is normalized according to
\begin{equation}
\vH \leftarrow {{10^{-\text{PG}_{\text{dB}}/20}}}{{\vH}}
\end{equation}
where $\text{PG}_{\text{dB}}$ is the path gain in
decibels.
Note that the information of the PG for each individual sample is contained in the channel
object generated with QuaDRiGa. Therefore, no further computation is required.

\begin{figure}[!t]
  \centering
\begin{tikzpicture}
\begin{axis}[
width = 9cm, height = 4.6cm,
tick align=outside,
tick pos=left,
x grid style={white!69.0196078431373!black},
xlabel={Carrier index},
xmin=-0.5, xmax=159.5,
xtick style={color=black},
xtick={-20,0,20,40,60,80,100,120,140,160},
xticklabels={\(\displaystyle -20\),\(\displaystyle 0\),\(\displaystyle 20\),\(\displaystyle 40\),\(\displaystyle 60\),\(\displaystyle 80\),\(\displaystyle 100\),\(\displaystyle 120\),\(\displaystyle 140\),\(\displaystyle 160\)},
y dir=reverse,
y grid style={white!69.0196078431373!black},
ylabel={Antenna index},
ymin=-0.5, ymax=63.5,
ytick style={color=black},
ytick={-20,0,20,40,60,80},
yticklabels={\(\displaystyle -20\),\(\displaystyle 0\),\(\displaystyle 20\),\(\displaystyle 40\),\(\displaystyle 60\),\(\displaystyle 80\)}
]
\addplot graphics [includegraphics cmd=\pgfimage,xmin=-0.5, xmax=159.5, ymin=63.5, ymax=-0.5] {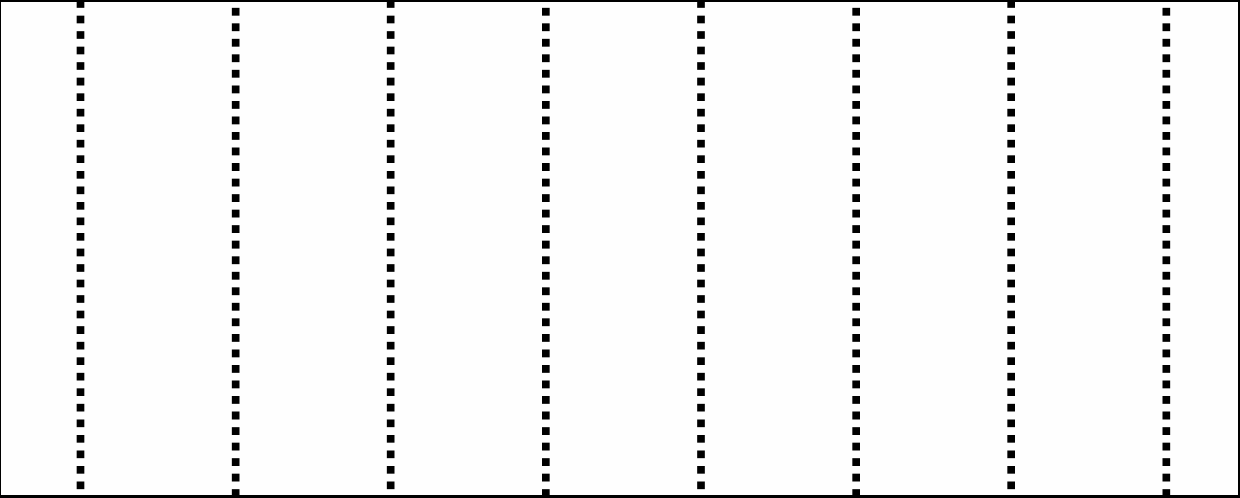};
\end{axis}

\end{tikzpicture}
  \caption{Example of binary mask with $ \eta = 0,025 $, where the black squares represent elements with value 1.}
  \label{fig: mask}
\end{figure}
\begin{table}[ht]
  \caption{Proposed CNN architecture.}
  \label{tab: cnn_lear}
  \begin{center}
  \begin{tabular}{lcccr}
  \hline
  Layer type& Output shape & \#Parameters $\boldsymbol{\theta} $\\
  \hline
  Input & $512$ & 0\\
  Reshape using mask & $64 \times 160 \times 2$ & 0\\
  Conv2D transposed, dilation=15 & $64 \times 160 \times 32$ & 608\\
  Batch normalization & $64\times 160\times 32$ & 128\\
  ReLU & $64\times 160\times 32$ & 0\\
  Conv2D transposed, dilation=7 & $64 \times 160 \times 64$ & 18496\\
  Batch normalization & $64\times 160\times 64$ & 256\\
  ReLU & $64\times 160\times 64$ & 0\\
  Conv2D transposed, dilation=4 & $64 \times 160 \times 128$ & 73856\\
  Batch normalization & $64\times 160\times 128$ & 512\\
  ReLU & $64\times 160\times 128$ & 0\\
  Conv2D transposed, dilation=2 & $64 \times 160 \times 64$ &  73792\\
  Batch normalization & $64\times 160\times 64$ & 256\\
  ReLU & $64\times 160\times 64$ & 0\\
  Conv2D transposed, dilation=2 & $64 \times 160 \times 32$ & 18464\\
  Batch normalization & $64\times 160\times 32$ & 128\\
  ReLU & $64\times 160\times 32$ & 0\\
  Conv2D transposed & $64 \times 160 \times 2$ & 578\\
  Multiplication (with reversed mask) & $64 \times 160 \times 2$ & 0\\
  Addition (with Input)& $64 \times 160 \times 2$ & 0\\
  \hline
\end{tabular}
\end{center}
\end{table}

\section{Learning in the uplink domain}
\label{sec: learn_ul}
\noindent In this section, the details of the training procedure are presented. As outlined in Section \ref{sec: intro}, we follow the conjecture that the reconstruction of the DL CSI based on a small portion of feedback information can be learned in the UL domain without any requirement of training data in the DL domain. Consequently, the training can be carried out entirely at the BS utilizing the UL CSI which is directly collected at the BS without any feedback requirements. Specifically, the UL channel matrix $\ulbig \in \mathbb{C}^{N_{\text{a}}\times N_{\text{c}}}$ is transformed into $\ulbig^{\text{real}} \in \mathbb{R}^{N_{\text{a}}\times
N_{\text{c}} \times 2}$ by stacking the real and imaginary parts along the third
dimension of a tensor. Thanks to this transformation, we work only with real-valued numbers. In the
next step, the UL channel matrix $\ulbig^\text{real}$ is downsampled to
\begin{equation}
  \ulpartial = \vectoriz(\ulbig^\text{real} \odot \vM) 
\end{equation}

\noindent where $\odot \ \vM$ represents a binary masking that only keeps a reduced amount of totally $2\eta N_{\text{a}} N_{\text{c}}$ real-valued entries of the channel matrix $\ulbig^\text{real}$ with $ 0 < \eta \ll 1 $, i.e., the \vari{compression ratio corresponding to the binary masking is given by $1/\eta$}.
In addition, $\vectoriz(\cdot)$ denotes the vectorization operation. The matrix $ \vM$ is designed
such that we just select out $2 \eta N_{\textit{c}}$ of all carriers and for each carrier we consider only half (every second) of the antennas coefficients. Moreover, the selected carriers are equidistantly spaced in the carrier domain. These choices of parameters are made with regard to the typically small size of affordable feedback information in the DL domain when analyzing the FDD system.

\noindent An example mask with $ \eta = 0.025 $ and thus a compression ratio of $40$ is illustrated in
Figure~\ref{fig: mask}, where $8$ out of $160$ carriers are considered, and for each carrier, $32$ out of $64$ antenna coefficients are selected. 
\vari{In order to attract the readers' attention to the method rather than to the optimization of the mask $\vM$, the rest of the paper is based on this simple mask. Further results with different masks can be found in Appendix~\ref{app: cae_results}.}

\begin{figure}[!t]
\vari{
\centering
  \subfloat[][Input.]{\label{fig: di-net-input}
  \includegraphics[scale=0.16]{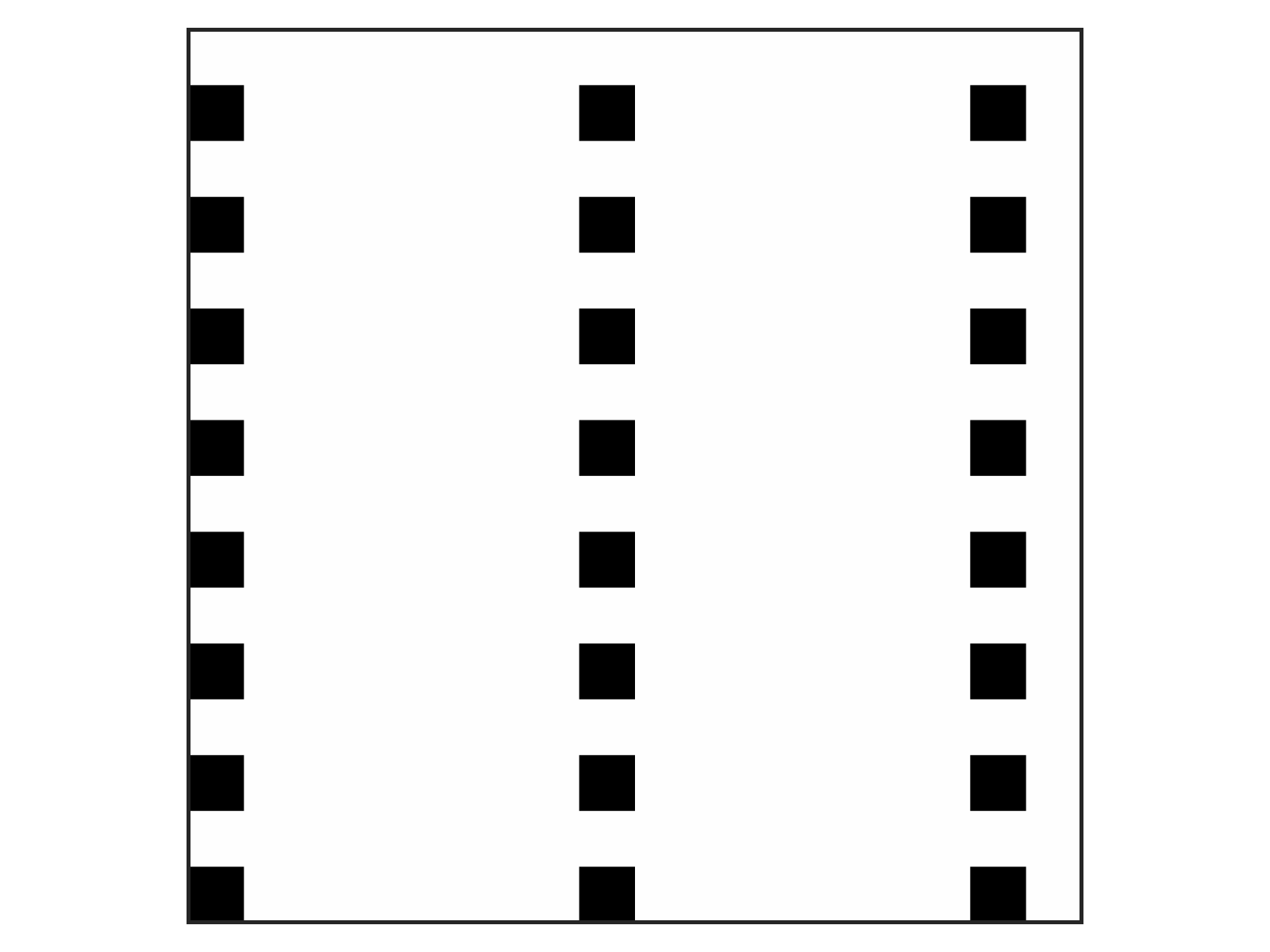}}\hspace{0.2cm}
  \subfloat[][Output without dilation.]{\label{fig: no-dil-out}
  \includegraphics[scale=0.16]{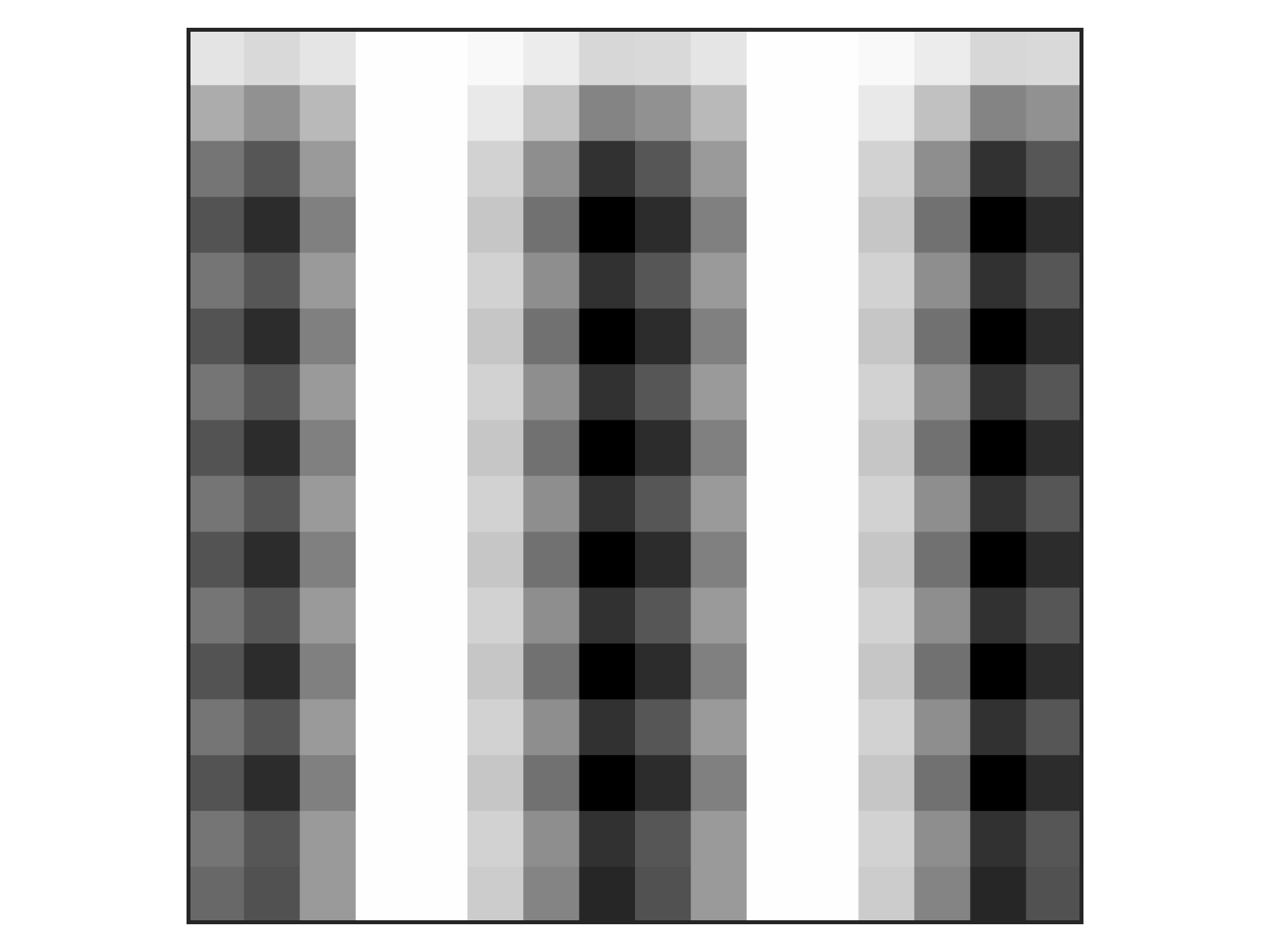}}\hspace{0.2cm}
  \subfloat[][Output with dilation.]{\label{fig: di-out}
  \includegraphics[scale=0.16]{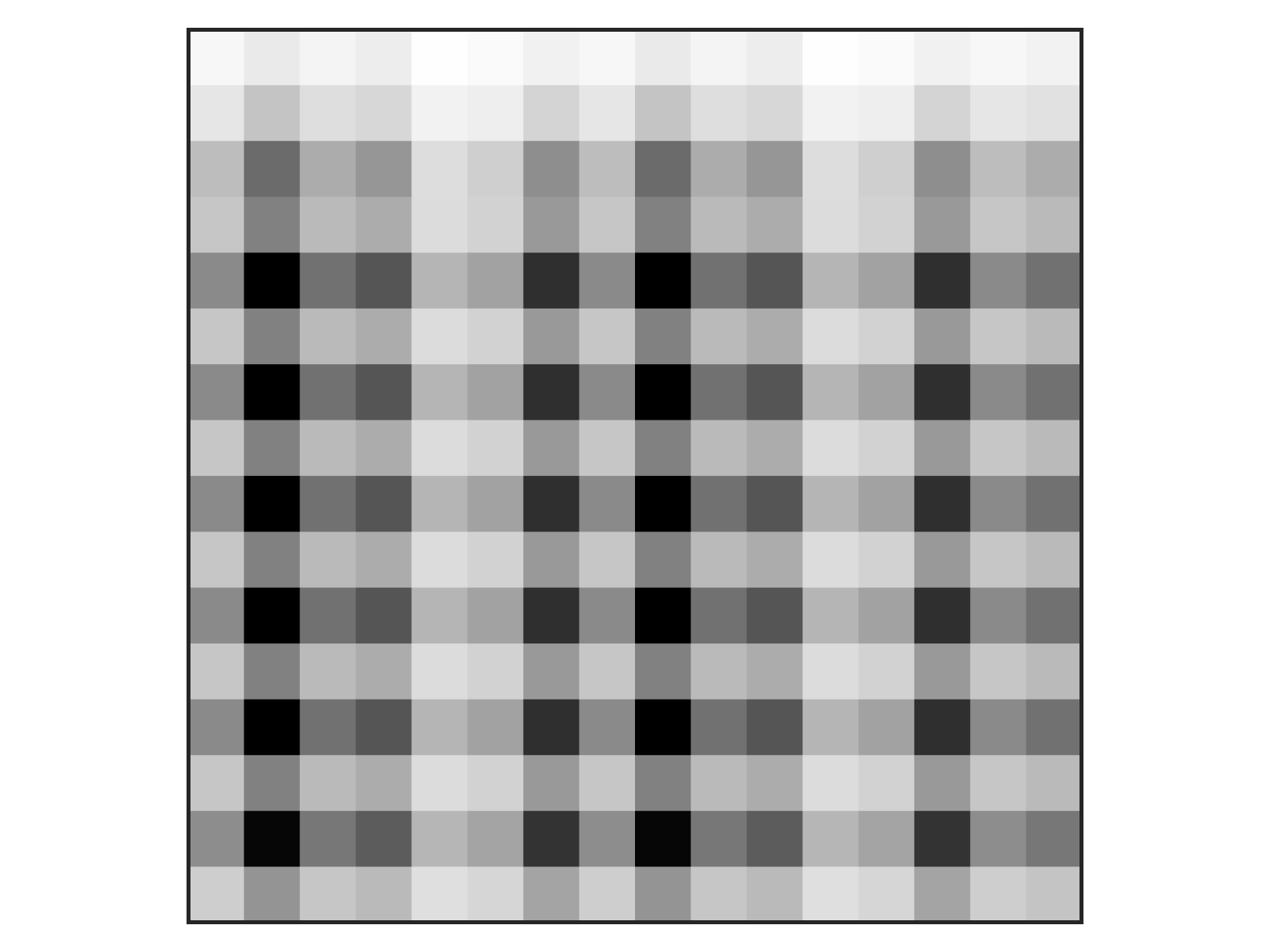}}
  \caption{Example of the affected output of a two layer convolutional neural network in each case with and without dilation. The ``pixels'' that are affected by the input are highlighted with gray levels. In both cases, the network is fed with the same binary sparse input, where the black squares represent a non-zero input value. It can be observed that using dilation larger than one in the first layer helps to progressively complete the matrix, compared to a neural network that uses standard convolutions.}}
\end{figure}

\noindent It follows a closer inspection of the CNN that instantiates the reconstruction function $f_\text{CNN}(\cdot;{\boldsymbol{\theta}})$ and the architectural details of the CNN are displayed in Table~\ref{tab: cnn_lear}. \vari{Note that the architecture is based on convolutional layers and that fully connected layers are \emph{avoided} in order to lower the number of training parameters. Moreover, convolutional architectures, once trained, are applicable to various channel dimensions as shown in~\cite{Mashhadi2021a}.  
In particular, here, except in the last layer, we always have convolutional layers with dilation larger then $1$, cf. \cite{YuKo15}. The advantage of using dilated convolutions for sparse inputs is illustrated in Figure~\ref{fig: di-net-input}--\ref{fig: di-out}, where the affected output in terms of matrix entries after two standard convolutional layers (Fig.~\ref{fig: no-dil-out}) is compared to the affected output obtained after two convolutional layers, where the first convolutional layer uses a dilatation of $2$ (Fig.~\ref{fig: di-out}). There, it can be observed that because of the dilation, the filters of the convolutional layers can be tuned in order to complete the full matrix, whereas with standard convolutions we have some residual error, since some matrix entries always remain unaffected, despite the filters we choose. One option for completing the matrix with standard convolutional layers could be to increase the number of layers itself, however, this would require different architectures for different sparsity levels.} 

\noindent Before feeding the input vector $\ulpartial$ in the convolutional layers, we reshape it as a sparse tensor with dimension
$N_{\text{a}} \times N_{\text{c}} \times 2$, where the elements of
$\ulpartial$ are located in the same places as the non-vanishing entries of the binary matrix $\vM$. After
each convolutional layer, the input tensor is gradually completed to eventually obtain
$\ulbighat^\text{real}$. The goal of the CNN is to instantiate a function $ f_\text{CNN}(\cdot;{\boldsymbol{\theta}}) $ which reconstructs an output $\ulbighat^\text{real}$ approximately equal to the original channel $\ulbig^\text{real}$, i.e.,
\begin{equation*}
  f_\text{CNN}(\ulpartial;{\boldsymbol{\theta}}) = \ulbighat^\text{real} \approxeq \ulbig^\text{real},
\end{equation*}
where $ \boldsymbol{\theta} $ refers to the adjustable weights of the CNN. To this end, the training phase of the CNN is based on a typical empirical risk function, the loss function of which is given by
\begin{align*}
  L(\boldsymbol{\theta},\ulbig^\text{real}) &= \norm { f_\text{CNN}(\vectoriz(\ulbig^\text{real} \odot \vM);{\boldsymbol{\theta}}) - \ulbig^\text{real}}^2.
\end{align*}

\section{Retrieval in the downlink domain}
\label{sec: dl_pred}
\noindent Once the parameters $\boldsymbol{\theta}$ of the CNN are learned, based only on UL data, the CNN is ``transfered'' to the DL frequency domain by applying it on the downsampled DL channel $ \dlfeedb $. In order to enable the reconstruction of the DL CSI, the downsampled DL channel $ \dlfeedb $ has to share the same formatting and size as $ \ulpartial $ in the UL domain. Consequently, the MT has to feed back the DL coefficients to the BS according to the non-vanishing entries of the mask $\vM$, 
i.e.,
\begin{equation}
\dlfeedb = \vectoriz(\dlbig^\text{real} \odot \vM).
\end{equation}
Subsequent to an equal reshaping of $\dlfeedb$ as introduced in Section~\ref{sec: learn_ul} for $\ulpartial$, the full DL CSI is eventually recovered by means of the trained CNN, i.e.,
\begin{equation}
  \dlbighat^\text{real} = f_\text{CNN}(\dlfeedb;{\boldsymbol{\theta}}).
\end{equation}
Note again, that the CNN applied to the DL data is based on the UL data.
\\
As already mentioned before, the basis for this straightforward application of the UL-trained CNN to the fed back DL CSI is the conjecture that learning the channel recovery in the UL domain is transferable to the DL domain. In Section~\ref{sec: verify_conj}, a statistical analysis based on the maximum
mean discrepancy (MMD) metric supports our argumentation.

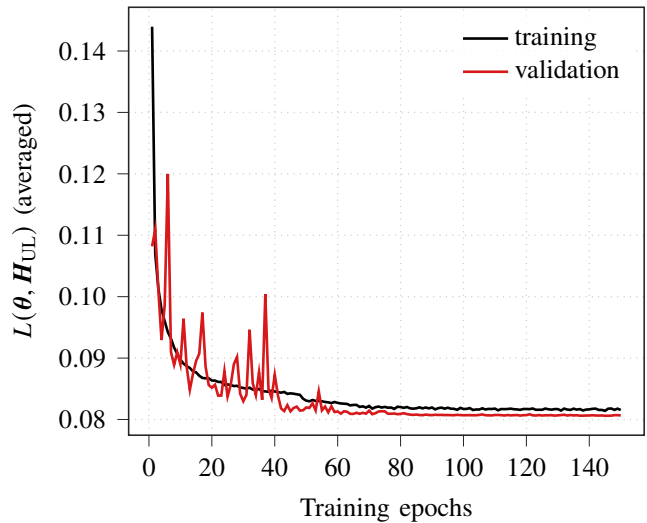
\begin{figure}[t!]
  \centering
\begin{tikzpicture}

\definecolor{color0}{HTML}{3498DB}
\definecolor{color1}{rgb}{0.843137254901961,0.0980392156862745,0.109803921568627}


\begin{axis}[
axis line style={white!15!black},
legend cell align={left},
legend style={fill opacity=0.8, draw opacity=1, text opacity=1, draw=none},
tick align=outside,
tick pos=left,
x grid style={white!80!black},
grid style = {dotted, color = black, line width = 0.5pt},
xlabel={Training epochs},
xmajorgrids,
xmin=-6.45, xmax=157.45,
xtick style={color=white!15!black},
xtick={-20,0,20,40,60,80,100,120,140,160},
xticklabels={\(\displaystyle -20\),\(\displaystyle 0\),\(\displaystyle 20\),\(\displaystyle 40\),\(\displaystyle 60\),\(\displaystyle 80\),\(\displaystyle 100\),\(\displaystyle 120\),\(\displaystyle 140\),\(\displaystyle 160\)},
y grid style={white!80!black},
ylabel={$L(\boldsymbol{\theta},\ulbig)$ (averaged)},
ymajorgrids,
ymin=0.0774408930289591, ymax=0.147118948031988,
ytick style={color=white!15!black},
ytick={0.07,0.08,0.09,0.1,0.11,0.12,0.13,0.14,0.15},
yticklabels={\(\displaystyle 0.07\),\(\displaystyle 0.08\),\(\displaystyle 0.09\),\(\displaystyle 0.10\),\(\displaystyle 0.11\),\(\displaystyle 0.12\),\(\displaystyle 0.13\),\(\displaystyle 0.14\),\(\displaystyle 0.15\)},
line width = 0.6pt
]
\addplot [line width=1pt, black]
table {%
1 0.143951763713668
2 0.10697737253128
3 0.101281704695247
4 0.0979080749599724
5 0.0960086828395491
6 0.0942451895439419
7 0.0931657902849692
8 0.0917154683249778
9 0.0906891381525694
10 0.0897625295381094
11 0.0891005625195077
12 0.0886951734781919
13 0.0883911312782559
14 0.0878301915540404
15 0.0876751818989044
16 0.0871230773055535
17 0.0867619896780548
18 0.0867103902570506
19 0.0867126734068962
20 0.0863453026347119
21 0.0863331799888779
22 0.086125181938834
23 0.0861075278383251
24 0.0856174426322634
25 0.0857069053503226
26 0.0856110796822836
27 0.0854441679231706
28 0.0854859505958131
29 0.0852266605254438
30 0.0851368525395386
31 0.085016268752165
32 0.0851780112587546
33 0.0848748323685697
34 0.0849357120260643
35 0.0849650524650061
36 0.0846060948043501
37 0.0845768599569517
38 0.0845136106609831
39 0.0846914116497557
40 0.0845629235793804
41 0.0844590063425815
42 0.084371677584082
43 0.0844871514047668
44 0.0841773182679213
45 0.0842612601734908
46 0.0841974717287528
47 0.0841064391967178
48 0.0840152901717021
49 0.0833824127725673
50 0.0830739498967568
51 0.082963176402132
52 0.0831719742416401
53 0.0830274143924814
54 0.0830062909688228
55 0.0829656682943569
56 0.0828589939328384
57 0.0827907316024681
58 0.0826745061758544
59 0.0828474242847448
60 0.0827087423667826
61 0.0826134688679281
62 0.0825737160855326
63 0.0825354747661042
64 0.0822989285377593
65 0.082373644736213
66 0.0823758226670442
67 0.0820476400429459
68 0.0821455740139215
69 0.0819369225783221
70 0.0822072632307169
71 0.0817679445781865
72 0.081999779739023
73 0.0821104113327859
74 0.0819448973387947
75 0.0820500858992245
76 0.0819207059937678
77 0.0819465663377488
78 0.0817230092495867
79 0.0820729160173373
80 0.082031384150156
81 0.0818988573063037
82 0.0820345888819347
83 0.0818233725080576
84 0.0817600112957267
85 0.0818234170060079
86 0.0819773696292904
87 0.0818136312861615
88 0.0819495560338598
89 0.0818143700696178
90 0.0819545936189755
91 0.0816457680727247
92 0.0818479392187068
93 0.081932073199777
94 0.0817671976749987
95 0.0819154803550636
96 0.0817597867255347
97 0.0817412013602574
98 0.0815671568777307
99 0.0818970993394762
100 0.0817121494386263
101 0.0817242873330428
102 0.0818596902604107
103 0.0815493444558875
104 0.0816321413520268
105 0.0816238640244104
106 0.0818188078000815
107 0.0816350391928725
108 0.0817287941839908
109 0.0817065178822481
110 0.0817739129323384
111 0.0814978686094191
112 0.0817019023713554
113 0.0816859721366701
114 0.0817050500682389
115 0.0817646856323398
116 0.0816743355991792
117 0.0816434975433219
118 0.0815737185199806
119 0.081706291307523
120 0.0817126369845344
121 0.0816881962108855
122 0.0817918255317921
123 0.081584629385519
124 0.081590156366926
125 0.0815888192819951
126 0.0817763203854284
127 0.0816673691533389
128 0.0816006490896488
129 0.0816959493574584
130 0.0816921682672161
131 0.0814559463610096
132 0.081776343176264
133 0.0816209056989807
134 0.0815579080306064
135 0.0817663695254483
136 0.0816082476249963
137 0.081697542345309
138 0.0814744715219755
139 0.0816193071664034
140 0.0816438514517392
141 0.0817379320875324
142 0.0817710734950525
143 0.081564162866589
144 0.0815592462950179
145 0.0814087459391281
146 0.0817642336857263
147 0.081787159755172
148 0.081567406772306
149 0.0817631918855222
150 0.0815484719695436
};
\addlegendentry{training}
\addplot [line width=1pt, color1]
table {%
1 0.108200751245022
2 0.110939279198647
3 0.101015895605087
4 0.0929267853498459
5 0.0985395163297653
6 0.119982406497002
7 0.0909696146845818
8 0.0888976976275444
9 0.0909333974123001
10 0.0889890193939209
11 0.0964110568165779
12 0.0885604545474052
13 0.0848341286182404
14 0.0874423980712891
15 0.08954057097435
16 0.0907217934727669
17 0.097407802939415
18 0.0886334627866745
19 0.0856258198618889
20 0.0851705968379974
21 0.0856549814343452
22 0.0838895440101624
23 0.0839012265205383
24 0.0878778919577599
25 0.0837295576930046
26 0.0856539383530617
27 0.088991716504097
28 0.0901746228337288
29 0.0841948762536049
30 0.0829745382070541
31 0.0840215682983398
32 0.0946172475814819
33 0.0859524980187416
34 0.0839656665921211
35 0.087672047317028
36 0.0832128599286079
37 0.100406937301159
38 0.0850079208612442
39 0.0834936201572418
40 0.087292917072773
41 0.0839415714144707
42 0.0818090662360191
43 0.0813743099570274
44 0.0823347643017769
45 0.0813290849328041
46 0.081878736615181
47 0.0820923373103142
48 0.0814561024308205
49 0.0814967304468155
50 0.0819576978683472
51 0.0818806663155556
52 0.0826463252305984
53 0.0814772620797157
54 0.0845741108059883
55 0.0814846605062485
56 0.0821075811982155
57 0.0811624452471733
58 0.0821520239114761
59 0.0812227129936218
60 0.0813107192516327
61 0.0809566825628281
62 0.081334725022316
63 0.0812584459781647
64 0.0809192210435867
65 0.0809146463871002
66 0.0810975283384323
67 0.0809259861707687
68 0.0810927748680115
69 0.0809477120637894
70 0.0814450904726982
71 0.0808626711368561
72 0.0810244232416153
73 0.0813008546829224
74 0.081337958574295
75 0.0813233554363251
76 0.0809332206845284
77 0.080943189561367
78 0.0809766575694084
79 0.0808640867471695
80 0.0809162929654121
81 0.0810272544622421
82 0.0808791369199753
83 0.0807991847395897
84 0.0807498767971992
85 0.0808037891983986
86 0.0807322338223457
87 0.0807027146220207
88 0.0808017626404762
89 0.0807315856218338
90 0.0807116106152534
91 0.0807695016264915
92 0.0807402282953262
93 0.0807531028985977
94 0.0807198882102966
95 0.0807263031601906
96 0.0807091146707535
97 0.0806916132569313
98 0.0807636752724648
99 0.0807162448763847
100 0.080692894756794
101 0.0807026699185371
102 0.0807019174098969
103 0.0806933045387268
104 0.0806901007890701
105 0.0807439908385277
106 0.0808276385068893
107 0.080694817006588
108 0.08067487180233
109 0.0806655660271645
110 0.0807340517640114
111 0.0807368233799934
112 0.0807083621621132
113 0.0806547254323959
114 0.0807293131947517
115 0.0806692615151405
116 0.0807025581598282
117 0.0807243883609772
118 0.0806383639574051
119 0.080700159072876
120 0.0807161033153534
121 0.0808538869023323
122 0.080682672560215
123 0.0806451067328453
124 0.0807173252105713
125 0.0806519314646721
126 0.0806853324174881
127 0.0806858316063881
128 0.0807052403688431
129 0.0806282982230186
130 0.0806801542639732
131 0.0807170867919922
132 0.0807763338088989
133 0.0806486457586288
134 0.0807221680879593
135 0.0806644931435585
136 0.080692820250988
137 0.0806286409497261
138 0.0806253254413605
139 0.0806526318192482
140 0.0806567370891571
141 0.0806910619139671
142 0.0806593596935272
143 0.0806408226490021
144 0.0806557461619377
145 0.0806080773472786
146 0.0806139186024666
147 0.0806280672550201
148 0.0807076916098595
149 0.0807090923190117
150 0.0806585252285004
};
\addlegendentry{validation}
\end{axis}

\end{tikzpicture}
  \caption{Evolution of training and validation loss during UL training.}
  \label{fig: loss}
\end{figure}

\section{Results and discussion}
\label{sec: result_single}
\noindent The CNN has been implemented with Tensorflow~\cite{tensorflow2015} and the
training has been done with single precision numbers. We consider mini-batches
of $64$ samples and we use the Adam optimizer~\cite{adam} to update the weights of the
neural network for every batch. An epoch consists of $3750$ batches and at the
end of each epoch we utilize the validation set only for computing the
validation loss. 
In
Figure~\ref{fig: loss}, it can be observed that $100$ epochs are required to reach convergence
in terms of validation loss, which corresponds to a number of batches in the
order of $10^6$. The typically high number of training examples in applications with CNNs poses little problem in the approach presented here, since the training data is based exclusively on UL CSI. It can be assumed that this is regularly estimated anyway during UL operation of the communication link and that a representative sampling of the propagation scenario is ensured over time due to the users being in the cell and moving around. A detailed study of such an acquisition of the radio properties of the environment is subject to further studies which are beyond the scope of this paper.

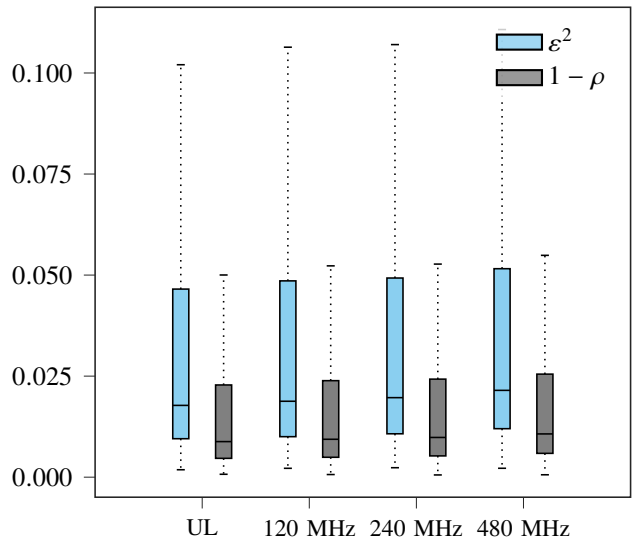
\begin{figure}[t!]
  \centering
\begin{tikzpicture}

\definecolor{color0}{rgb}{0.54, 0.81, 0.94}

\begin{axis}[
width = 8.7cm,
height = 8.1cm,
axis line style={white!15!black},
legend cell align={left},
legend style={fill opacity=0.8, draw opacity=1, text opacity=1, draw=none},
tick align=outside,
tick pos=left,
x grid style={white!80!black},
xmin=-2, xmax=8,
xtick style={color=white!15!black},
xtick={0,2,4,6},
xticklabels={{\footnotesize{UL}}, {\footnotesize{$120$~MHz}},{\footnotesize{$240$~MHz}},{\footnotesize{$480$~MHz}}},
xticklabel style   = {align=center},
y grid style={white!80!black},
ymin=-0.00495739802718163, ymax=0.116259818524122,
ytick style={color=white!15!black},
ytick={-0.025,0,0.025,0.05,0.075,0.1,0.125},
yticklabels={\(\displaystyle -0.025\),\(\displaystyle 0.000\),\(\displaystyle 0.025\),\(\displaystyle 0.050\),\(\displaystyle 0.075\),\(\displaystyle 0.100\),\(\displaystyle 0.125\)},
line width = 0.6pt
]
\addlegendimage{area legend, thick,fill = color0}
\addlegendentry{$\varepsilon^2$}

\addlegendimage{area legend, thick,fill = gray}
\addlegendentry{$1-\rho$}

\addplot [black, dotted, forget plot]
table {%
-0.4 0.00951476255431771
-0.4 0.00183793250471354
};
\addplot [black, dotted, forget plot]
table {%
-0.4 0.0465349713340402
-0.4 0.102042317390442
};
\addplot [black, forget plot]
table {%
-0.475 0.00183793250471354
-0.325 0.00183793250471354
};
\addplot [black, forget plot]
table {%
-0.475 0.102042317390442
-0.325 0.102042317390442
};
\addplot [black, dotted, forget plot]
table {%
1.6 0.010019056731835
1.6 0.00219337688758969
};
\addplot [black, dotted, forget plot]
table {%
1.6 0.0485677244141698
1.6 0.106382824480534
};
\addplot [black, forget plot]
table {%
1.525 0.00219337688758969
1.675 0.00219337688758969
};
\addplot [black, forget plot]
table {%
1.525 0.106382824480534
1.675 0.106382824480534
};
\addplot [black, dotted, forget plot]
table {%
3.6 0.0107339585665613
3.6 0.00233903457410634
};
\addplot [black, dotted, forget plot]
table {%
3.6 0.0492726620286703
3.6 0.107024423778057
};
\addplot [black, forget plot]
table {%
3.525 0.00233903457410634
3.675 0.00233903457410634
};
\addplot [black, forget plot]
table {%
3.525 0.107024423778057
3.675 0.107024423778057
};
\addplot [black, dotted, forget plot]
table {%
5.6 0.0119963060133159
5.6 0.00220230012200773
};
\addplot [black, dotted, forget plot]
table {%
5.6 0.0515772746875882
5.6 0.110749945044518
};
\addplot [black, forget plot]
table {%
5.525 0.00220230012200773
5.675 0.00220230012200773
};
\addplot [black, forget plot]
table {%
5.525 0.110749945044518
5.675 0.110749945044518
};
\addplot [black, dotted, forget plot]
table {%
0.4 0.00466032326221466
0.4 0.000716567039489746
};
\addplot [black, dotted, forget plot]
table {%
0.4 0.022814154624939
0.4 0.0500226020812988
};
\addplot [black, forget plot]
table {%
0.325 0.000716567039489746
0.475 0.000716567039489746
};
\addplot [black, forget plot]
table {%
0.325 0.0500226020812988
0.475 0.0500226020812988
};
\addplot [black, dotted, forget plot]
table {%
2.4 0.00492340326309204
2.4 0.000651538372039795
};
\addplot [black, dotted, forget plot]
table {%
2.4 0.0238783955574036
2.4 0.0522812604904175
};
\addplot [black, forget plot]
table {%
2.325 0.000651538372039795
2.475 0.000651538372039795
};
\addplot [black, forget plot]
table {%
2.325 0.0522812604904175
2.475 0.0522812604904175
};
\addplot [black, dotted, forget plot]
table {%
4.4 0.00526228547096252
4.4 0.000552475452423096
};
\addplot [black, dotted, forget plot]
table {%
4.4 0.0242573469877243
4.4 0.0527189373970032
};
\addplot [black, forget plot]
table {%
4.325 0.000552475452423096
4.475 0.000552475452423096
};
\addplot [black, forget plot]
table {%
4.325 0.0527189373970032
4.475 0.0527189373970032
};
\addplot [black, dotted, forget plot]
table {%
6.4 0.00589138269424438
6.4 0.000592827796936035
};
\addplot [black, dotted, forget plot]
table {%
6.4 0.0254949182271957
6.4 0.054883599281311
};
\addplot [black, forget plot]
table {%
6.325 0.000592827796936035
6.475 0.000592827796936035
};
\addplot [black, forget plot]
table {%
6.325 0.054883599281311
6.475 0.054883599281311
};
\path [draw=black, fill=color0]
(axis cs:-0.55,0.00951476255431771)
--(axis cs:-0.25,0.00951476255431771)
--(axis cs:-0.25,0.0465349713340402)
--(axis cs:-0.55,0.0465349713340402)
--(axis cs:-0.55,0.00951476255431771)
--cycle;
\path [draw=black, fill=color0]
(axis cs:1.45,0.010019056731835)
--(axis cs:1.75,0.010019056731835)
--(axis cs:1.75,0.0485677244141698)
--(axis cs:1.45,0.0485677244141698)
--(axis cs:1.45,0.010019056731835)
--cycle;
\path [draw=black, fill=color0]
(axis cs:3.45,0.0107339585665613)
--(axis cs:3.75,0.0107339585665613)
--(axis cs:3.75,0.0492726620286703)
--(axis cs:3.45,0.0492726620286703)
--(axis cs:3.45,0.0107339585665613)
--cycle;
\path [draw=black, fill=color0]
(axis cs:5.45,0.0119963060133159)
--(axis cs:5.75,0.0119963060133159)
--(axis cs:5.75,0.0515772746875882)
--(axis cs:5.45,0.0515772746875882)
--(axis cs:5.45,0.0119963060133159)
--cycle;
\path [draw=black, fill=gray]
(axis cs:0.25,0.00466032326221466)
--(axis cs:0.55,0.00466032326221466)
--(axis cs:0.55,0.022814154624939)
--(axis cs:0.25,0.022814154624939)
--(axis cs:0.25,0.00466032326221466)
--cycle;
\path [draw=black, fill=gray]
(axis cs:2.25,0.00492340326309204)
--(axis cs:2.55,0.00492340326309204)
--(axis cs:2.55,0.0238783955574036)
--(axis cs:2.25,0.0238783955574036)
--(axis cs:2.25,0.00492340326309204)
--cycle;
\path [draw=black, fill=gray]
(axis cs:4.25,0.00526228547096252)
--(axis cs:4.55,0.00526228547096252)
--(axis cs:4.55,0.0242573469877243)
--(axis cs:4.25,0.0242573469877243)
--(axis cs:4.25,0.00526228547096252)
--cycle;
\path [draw=black, fill=gray]
(axis cs:6.25,0.00589138269424438)
--(axis cs:6.55,0.00589138269424438)
--(axis cs:6.55,0.0254949182271957)
--(axis cs:6.25,0.0254949182271957)
--(axis cs:6.25,0.00589138269424438)
--cycle;
\addplot [black, forget plot]
table {%
-0.55 0.0177599005401134
-0.25 0.0177599005401134
};
\addplot [black, forget plot]
table {%
1.45 0.0187821295112371
1.75 0.0187821295112371
};
\addplot [black, forget plot]
table {%
3.45 0.0196741865947843
3.75 0.0196741865947843
};
\addplot [black, forget plot]
table {%
5.45 0.0214795991778374
5.75 0.0214795991778374
};
\addplot [black, forget plot]
table {%
0.25 0.00879955291748047
0.55 0.00879955291748047
};
\addplot [black, forget plot]
table {%
2.25 0.00937563180923462
2.55 0.00937563180923462
};
\addplot [black, forget plot]
table {%
4.25 0.00982704758644104
4.55 0.00982704758644104
};
\addplot [black, forget plot]
table {%
6.25 0.0107007324695587
6.55 0.0107007324695587
};
\end{axis}

\end{tikzpicture}
  \caption{NMSE and cosine similarity for recovered UL CSI and DL CSI based on the same CNN and propagation scenario.}
  \label{fig: metrics_same_cell}
\end{figure}

\begin{figure*}[t!]
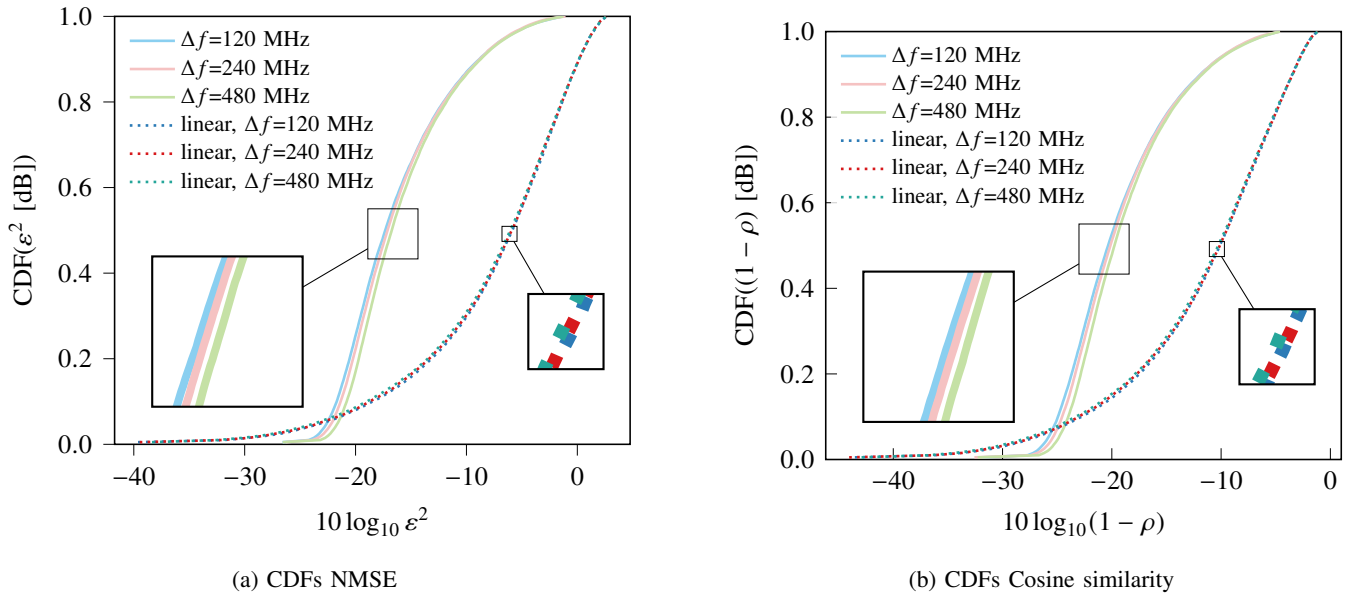

  \subfloat[][CDFs NMSE]{\label{fig: cdf nmse}
  \input{nmse_new.tex}}\hspace{1cm} 
  \subfloat[][CDFs Cosine similarity]{\label{fig: cdf cossim}\input{rho_new.tex}}
  \caption{Empirical CDFs of NMSE and cosine similarity for recovered UL and DL based on the same CNN and propagation scenario. 
  }
\end{figure*}

\noindent After the training, the performance is evaluated in terms of normalized mean square error $ \varepsilon^2 $ and cosine similarity $\rho$ for the test sets with frequency gaps of $\Delta f = 120...480$ MHz,
where
\begin{equation}
  \varepsilon^2 = \mathbb{E}\left[\frac{\norm{\hat{\vect{H}} - \vect{H}}_{\mathrm{F}}^2}{\norm{\vect{H}}_{\mathrm{F}}^2}\right],
\end{equation}
and
\begin{equation}
\rho = \mathbb{E}\left[\frac{1}{N_{\text{c}}}\sum_{n=1}^{N_{\text{c}}}\frac{\vert\hat{\vect{h}}^{\text{H}}_n\vect{h}_n\vert}{\norm{\hat{\vect{h}}_n}_2 \norm{\vect{h}_n}_2}\right],
\end{equation}
with $\vect{H} \in \mathbb{C}^{N_{\text{a}}\times N_{\text{c}}}$ and its $n$-th column ${\vect{h}}_n$, and $\hat{\vect{H}}$ and
$\hat{\vect{h}}_n$ their corresponding reconstructed versions. The results of the performance metrics are shown
in Figure~\ref{fig: metrics_same_cell}, where for each box the median, the first quartile ($Q_1$) and third quartile ($Q_3$) are highlighted. \vari{The whiskers in the box-plots are chosen to be one and half time of the interquartile range ($\operatorname{IQR}$) below the first quartile, and above the third quartile (in formulas, $Q_1 - 1.5\times \operatorname{IQR}$ and $Q_3 + 1.5\times \operatorname{IQR}$). Note that $Q_1 $ represents the $25$th percentile of the data, $Q_3$ represents the $75$th percentile of the data, and $\operatorname{IQR}$ represents the difference between the third and the first quartile. The values outside the range covered by the whiskers are considered outliers.}\\
In Figures~\ref{fig: cdf nmse} and~\ref{fig: cdf cossim}, the CDFs of $ \varepsilon^2 $ and
$\rho$ are additionally displayed on a logarithmic scale and compared with the retrieved DL CSI recoveries with a conventional linear interpolation of the fed back CSI in the frequency domain \cite{1033876}. The latter serves as a reference in this work.
It is clear from Figure \ref{fig: metrics_same_cell} and Figures \ref{fig: cdf nmse} and~\ref{fig: cdf cossim} that the CNN for DL CSI performs very well, with only a slight drop in performance for increasing frequency spacing, even though the CNN has never seen training samples from the downlink frequency domain and the beamsquint effect for different center frequencies has been taken into account. For all scenarios, the uplink-based CNN clearly outperforms the reference solution. For other compression ratios $ \eta $ the reader may refer to Appendix~\ref{app: larger_cr} to obtain a first impression. A detailed study of the appropriate compression technique and an investigation of the trade-off between compression ratios and achievable performance is considered to be subjects of future research.

\subsection{Sum rate results and discussion}
\label{subsec: rate_res}
\noindent Although mean square error and cosine similarity are well-known and established performance criteria, this section additionally evaluates the quality of the channel reconstruction in a multi-user communication scenario, which in contrast to a single-user scenario is more susceptible to inaccurate CSI. In particular, we examine the quality of the channel reconstructions in terms of their achievable sum rate in a multi-user downlink scenario. To this end, the Linear Successive Allocation Algorithm (LISA)~\cite{GuUtDi09b,lisa} is applied to randomly selected test channels.\footnote{Please note that any other multi-user precoding technique can be applied as well.} Each channel is associated with a channel matrix in the DL frequency domain, which is then subject to feedback and the subsequent machine learning based recovery at the BS. LISA is a zero-forcing based precoding technique that simultaneously performs combined data stream and user selection, resulting in an LQ decomposition of the overall channel matrix consisting of the selected subchannels. The lower triangular matrix (L-factor) of the decomposition corresponds to the effective channel of the resulting pre-equalized system. If elaborate nonlinear coding schemes are not an option \cite{1705034}, a second precoding step transforms the resulting channel diagonally and results in a zero-forcing solution of the selected users. Finally, a water filling procedure is applied to the diagonal channel. For simplicity, the linear version of LISA is applied independently on each of the $160$ carriers of the communication links and the results are then averaged over the carriers.

\noindent Furthermore, we considered four different scenarios with $1$, $2$, $4$, and $8$ users. 

\noindent Figures~\ref{fig: lisa a}--\ref{fig: lisa c} 
show the average achievable per-user rate for a $120...480$~MHz frequency gap over $100$ instances of LISA simulation runs, respectively. The continuous lines represent the rate achievable with perfect DL CSI knowledge, the dashed lines represent the rate obtained with the DL CSI predicted with the CNN, and the dotted lines represent the rate achievable with the DL CSI as result of the linear interpolation of the DL feedback. We can observe that the achievable rate per user with the channels recovered with the CNN is close to the case of perfect CSI and that the gain of the CNN approach compared to the linear interpolation method is particularly salient in a multi-user setup, while in single-user scenarios the linear interpolation technique is sufficient. This is due to the known lower CSI requirements in cases where preequalization of multiple channels is not required. However, the linear interpolation technique clearly fails in multi-user scenarios.

\section{Applying the CNN in unknown environments}
\label{sec: cnn_on_another_cell}

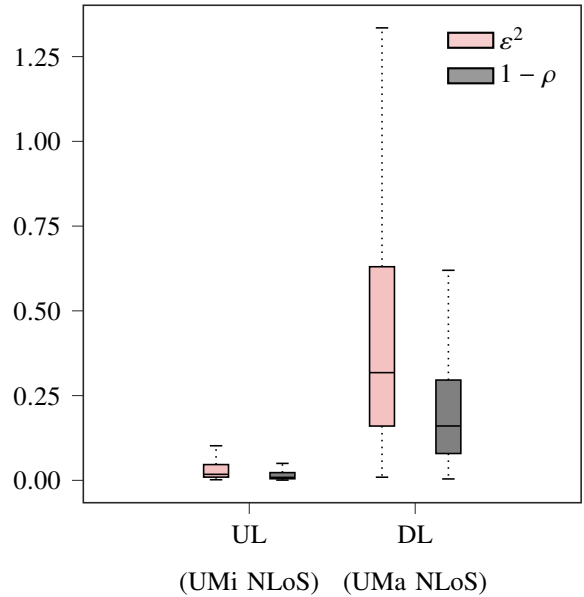
\begin{figure}[t!]
  \centering
\begin{tikzpicture}

\definecolor{color0}{rgb}{0.96, 0.76, 0.76}


\begin{axis}[
width = 8.2cm,
height = 8.2cm,
axis line style={white!15!black},
legend cell align={left},
legend style={fill opacity=0.8, draw opacity=1, text opacity=1, draw=none},
tick align=outside,
tick pos=left,
x grid style={white!80!black},
xmin=-2, xmax=4,
xtick style={color=white!15!black},
xtick={0,2},
xticklabels={{\small{UL}\\\small{(UMi NLoS)}}, {\small{DL}\\\small{(UMa NLoS)}}},
xticklabel style   = {align=center},
y grid style={white!80!black},
ymin=-0.0659848630428314, ymax=1.40144659876823,
ytick style={color=white!15!black},
ytick={-0.25,0,0.25,0.5,0.75,1,1.25,1.5},
yticklabels={\(\displaystyle -0.25\),\(\displaystyle 0.00\),\(\displaystyle 0.25\),\(\displaystyle 0.50\),\(\displaystyle 0.75\),\(\displaystyle 1.00\),\(\displaystyle 1.25\),\(\displaystyle 1.50\)},
line width = 0.6pt
]
\addlegendimage{area legend, thick,fill = color0}
\addlegendentry{$\varepsilon^2$}

\addlegendimage{area legend, thick,fill = gray}
\addlegendentry{$1-\rho$}
\addplot [black, dotted, forget plot]
table {%
-0.4 0.00951476255431771
-0.4 0.00183793250471354
};
\addplot [black, dotted, forget plot]
table {%
-0.4 0.0465349713340402
-0.4 0.102042317390442
};
\addplot [black, forget plot]
table {%
-0.475 0.00183793250471354
-0.325 0.00183793250471354
};
\addplot [black, forget plot]
table {%
-0.475 0.102042317390442
-0.325 0.102042317390442
};
\addplot [black, dotted, forget plot]
table {%
1.6 0.160149607807398
1.6 0.00927293766289949
};
\addplot [black, dotted, forget plot]
table {%
1.6 0.630166217684746
1.6 1.33474516868591
};
\addplot [black, forget plot]
table {%
1.525 0.00927293766289949
1.675 0.00927293766289949
};
\addplot [black, forget plot]
table {%
1.525 1.33474516868591
1.675 1.33474516868591
};
\addplot [black, dotted, forget plot]
table {%
0.4 0.00466032326221466
0.4 0.000716567039489746
};
\addplot [black, dotted, forget plot]
table {%
0.4 0.022814154624939
0.4 0.0500226020812988
};
\addplot [black, forget plot]
table {%
0.325 0.000716567039489746
0.475 0.000716567039489746
};
\addplot [black, forget plot]
table {%
0.325 0.0500226020812988
0.475 0.0500226020812988
};
\addplot [black, dotted, forget plot]
table {%
2.4 0.0791543275117874
2.4 0.00415313243865967
};
\addplot [black, dotted, forget plot]
table {%
2.4 0.295894458889961
2.4 0.619619071483612
};
\addplot [black, forget plot]
table {%
2.325 0.00415313243865967
2.475 0.00415313243865967
};
\addplot [black, forget plot]
table {%
2.325 0.619619071483612
2.475 0.619619071483612
};
\path [draw=black, fill=color0]
(axis cs:-0.55,0.00951476255431771)
--(axis cs:-0.25,0.00951476255431771)
--(axis cs:-0.25,0.0465349713340402)
--(axis cs:-0.55,0.0465349713340402)
--(axis cs:-0.55,0.00951476255431771)
--cycle;
\path [draw=black, fill=color0]
(axis cs:1.45,0.160149607807398)
--(axis cs:1.75,0.160149607807398)
--(axis cs:1.75,0.630166217684746)
--(axis cs:1.45,0.630166217684746)
--(axis cs:1.45,0.160149607807398)
--cycle;
\path [draw=black, fill=gray]
(axis cs:0.25,0.00466032326221466)
--(axis cs:0.55,0.00466032326221466)
--(axis cs:0.55,0.022814154624939)
--(axis cs:0.25,0.022814154624939)
--(axis cs:0.25,0.00466032326221466)
--cycle;
\path [draw=black, fill=gray]
(axis cs:2.25,0.0791543275117874)
--(axis cs:2.55,0.0791543275117874)
--(axis cs:2.55,0.295894458889961)
--(axis cs:2.25,0.295894458889961)
--(axis cs:2.25,0.0791543275117874)
--cycle;
\addplot [black, forget plot]
table {%
-0.55 0.0177599005401134
-0.25 0.0177599005401134
};
\addplot [black, forget plot]
table {%
1.45 0.31774839758873
1.75 0.31774839758873
};
\addplot [black, forget plot]
table {%
0.25 0.00879955291748047
0.55 0.00879955291748047
};
\addplot [black, forget plot]
table {%
2.25 0.160431414842606
2.55 0.160431414842606
};
\end{axis}

\end{tikzpicture}
  \caption{NMSE and cosine similarity for UL CSI versus DL CSI.}
  \label{fig: metrics_othe_cell}
\end{figure}

\noindent In this section, to investigate the scenario dependency, we apply the trained CNN
based on the UL CSI to the DL feedback of another cell with different
properties which were unknown during the training of the CNN. In particular, we considered the urban macrocell (UMa) NLoS scenario from QuaDRiGa. The
only elements in common between the cell used for the UL training and the other  cell are number and the type of
antennas considered at the BS and the MS, the bandwidth, and the number of
carriers considered. \vari{Specifically, the number of MPCs in UMi NLoS and UMa NLoS are relatively similar (61 in UMi and 64 in UMa). However, the UMa and UMi scenarios
differ not only in the number of MPCs, but in many other parameters involved in the
channel realization.}\\
The results of NMSE and the cosine similarity are shown in Figure~\ref{fig: metrics_othe_cell}. When we compare the performance metrics of the UMa NLoS cell with those of the UL baseline we can see
that the CNN, which was trained on a UMi NLoS scenario does not generalize well to the UMa NLoS scenario.

This is due to the model-free learning of the CNN, which is based on the specific UL data of the cell and its propagation characteristics that is different from the data of the other cell to which the CNN is applied. 
However, further research, which we do not present here, has revealed that the CNN can also generalize to scenarios with similar characteristics to the scenario on which the CNN was trained, or to scenarios in which the CSI is frequency flatter. Nevertheless, the results in this section show that in principle we cannot apply the CNN trained on one propagation environment to any other arbitrary environment without extra measures.
\vari{Please, note that considering changes in the environment
is beyond the scope of this specific work. Regarding the actual implementation
of the proposed architecture one could expect a three-phases approach similar to the one in~\cite{Mashhadi2021} with the key-difference that in our case the training phase is \textit{centralized} at the BS rather than \textit{federated} over multiple devices. Moreover, since the training is carried with the
sole UL data, the BS might update the weights of the CNN regularly as background
activity. In this way, the algorithm can be robust to changes in the environment, also taking into account that severe changes in the environment, as the construction of a new building, happen over a long time interval.}
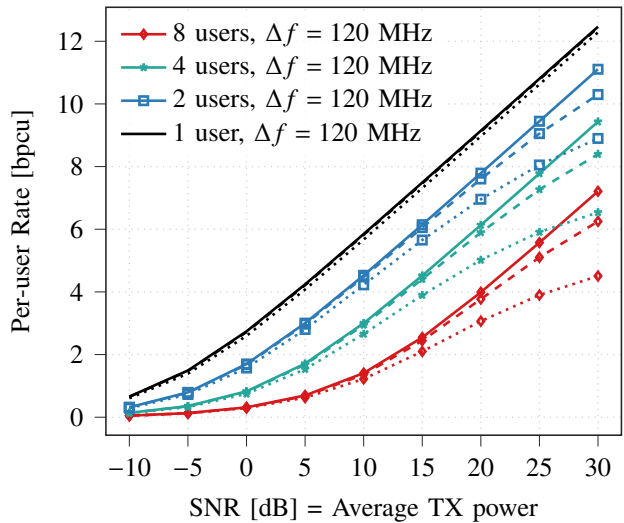
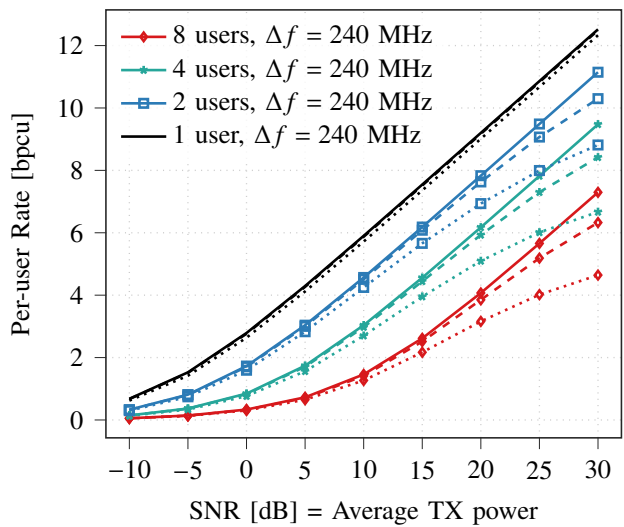
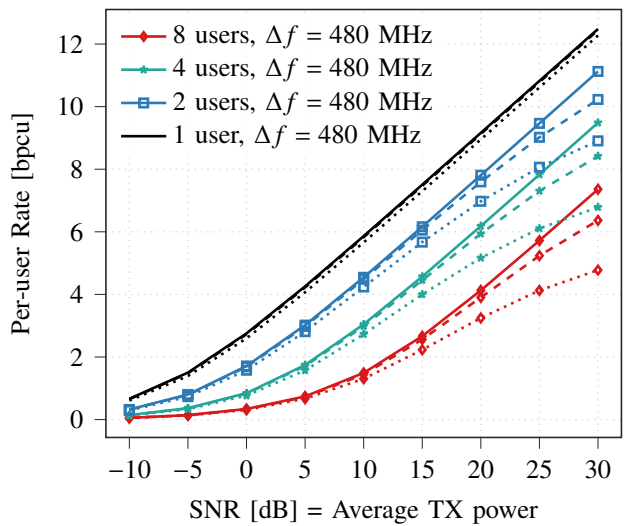
\begin{figure}[!p]
\centering
  \subfloat[][$\Delta f=120$~MHz]{\label{fig: lisa
  a}
\begin{tikzpicture}

\definecolor{strongred}{rgb}{0.843137254901961,0.0980392156862745,0.109803921568627}
\definecolor{strongblue}{rgb}{0.172549019607843,0.482352941176471,0.713725490196078}
\definecolor{babyblue}{rgb}{0.54, 0.81, 0.94}
\definecolor{babypink}{rgb}{0.96, 0.76, 0.76}
\definecolor{babygreen}{HTML}{C5E1A5}
\definecolor{stronggreen}{HTML}{26A69A}

\begin{axis}[
axis line style={white!15!black},
legend cell align={left},
legend style={fill opacity=0.8, draw opacity=1, text opacity=1, at={(0.01,0.99)}, anchor=north west, draw=none},
tick align=outside,
tick pos=left,
x grid style={white!80!black},
grid style = {dotted, color = black, line width = 0.5pt},
xlabel={SNR [dB] = Average TX power},
xmajorgrids,
xmin=-12, xmax=32,
xtick style={color=white!15!black},
xtick={-15,-10,-5,0,5,10,15,20,25,30,35},
xticklabels={\(\displaystyle -15\),\(\displaystyle -10\),\(\displaystyle -5\),\(\displaystyle 0\),\(\displaystyle 5\),\(\displaystyle 10\),\(\displaystyle 15\),\(\displaystyle 20\),\(\displaystyle 25\),\(\displaystyle 30\),\(\displaystyle 35\)},
y grid style={white!80!black},
ylabel={Per-user Rate [bpcu]},
ymajorgrids,
ymin=-0.574210738274809, ymax=13.0849237455529,
ytick style={color=white!15!black},
ytick={-2,0,2,4,6,8,10,12,14},
yticklabels={\(\displaystyle -2\),\(\displaystyle 0\),\(\displaystyle 2\),\(\displaystyle 4\),\(\displaystyle 6\),\(\displaystyle 8\),\(\displaystyle 10\),\(\displaystyle 12\),\(\displaystyle 14\)},
line width = 0.6pt
]
\addplot [line width=1pt, strongred, mark=diamond, mark size=1.6, mark options={solid,fill opacity=0}]
table {%
-10 0.0498816480251544
-5 0.131348004055268
0 0.315298093009574
5 0.694898548267725
10 1.4090482023328
15 2.5462619131671
20 3.98969689587832
25 5.57410648013217
30 7.20994871752298
};
\addlegendentry{8 users, $\Delta f=120$~MHz}
\addplot [line width=1pt, strongred, dashed, mark=diamond, mark size=1.6, mark options={solid,fill opacity=0}, forget plot]
table {%
-10 0.0484518095972236
-5 0.12776457370039
0 0.306618276725533
5 0.67490722661985
10 1.36228193379584
15 2.44339256605564
20 3.76962479344706
25 5.10362295631146
30 6.25041017346746
};
\addplot [line width=1pt, strongred, dotted, mark=diamond, mark size=1.6, mark options={solid,fill opacity=0}, forget plot]
table {%
-10 0.046659010990087
-5 0.122389016050652
0 0.290508561468624
5 0.624730117584143
10 1.221515075077
15 2.09944119309877
20 3.06605954563205
25 3.89991547316133
30 4.50651908057994
};
\addplot [line width=1pt, stronggreen, mark=star, mark size=1.6, mark options={solid,fill opacity=0}]
table {%
-10 0.138887290273236
-5 0.35469832819295
0 0.826278027285868
5 1.71125542003189
10 3.00044740924636
15 4.51999855667543
20 6.13309357882538
25 7.77857134675395
30 9.43460212203373
};
\addlegendentry{4 users, $\Delta f=120$~MHz}
\addplot [line width=1pt, stronggreen, dashed, mark=star, mark size=1.6, mark options={solid,fill opacity=0}, forget plot]
table {%
-10 0.13642760361737
-5 0.348443851359343
0 0.811073110131659
5 1.67996139406283
10 2.94283063749356
15 4.40814190148335
20 5.89608405492766
25 7.26806418285682
30 8.39538103923284
};
\addplot [line width=1pt, stronggreen, dotted, mark=star, mark size=1.6, mark options={solid,fill opacity=0}, forget plot]
table {%
-10 0.125696138989888
-5 0.322038245177571
0 0.746350978235532
5 1.53213443829102
10 2.65754325752581
15 3.89505960379447
20 5.0192675770394
25 5.90863702989959
30 6.53848918184217
};
\addplot [line width=1pt, strongblue, mark=square, mark size=1.6, mark options={solid,fill opacity=0}]
table {%
-10 0.317865663367582
-5 0.792948303186603
0 1.69906249029474
5 3.00310817901316
10 4.52910280991467
15 6.14448393916971
20 7.79071286493413
25 9.44698412186909
30 11.1064608294433
};
\addlegendentry{2 users, $\Delta f=120$~MHz}
\addplot [line width=1pt, strongblue, dashed, mark=square, mark size=1.6, mark options={solid,fill opacity=0}, forget plot]
table {%
-10 0.313228968121421
-5 0.781836326416759
0 1.67781111250998
5 2.96987706639614
10 4.47749285959063
15 6.05246111925741
20 7.604058120939
25 9.051858004332
30 10.2990566370326
};
\addplot [line width=1pt, strongblue, dotted, mark=square, mark size=1.6, mark options={solid,fill opacity=0}, forget plot]
table {%
-10 0.288502127809084
-5 0.726483158027236
0 1.57127064816098
5 2.802009629089
10 4.22612361528503
15 5.65481076912233
20 6.95722883047754
25 8.05062563229943
30 8.89875186574233
};
\addplot [line width=1pt, black]
table {%
-10 0.658353001470153
-5 1.49059062572314
0 2.73997117412348
5 4.24025052651623
10 5.84617053117094
15 7.48924952313155
20 9.14450791802221
25 10.8036626216061
30 12.464053996288
};
\addlegendentry{1 user, $\Delta f=120$~MHz}
\addplot [line width=1pt, black, dashed, forget plot]
table {%
-10 0.650157478059565
-5 1.47539700602963
0 2.71871719960517
5 4.21564922253537
10 5.8201899227386
15 7.46278303615823
20 9.11788173709521
25 10.7769852905354
30 12.4373604234474
};
\addplot [line width=1pt, black, dotted, forget plot]
table {%
-10 0.604881224382756
-5 1.3899116900806
0 2.59751523540757
5 4.07463027429041
10 5.67101811006982
15 7.31072024552747
20 8.96486148658797
25 10.6236571536964
30 12.2839343778708
};
\end{axis}

\end{tikzpicture}}\\
  \subfloat[][$\Delta f=240$~MHz]{\label{fig: lisa
  b}
\begin{tikzpicture}

\definecolor{strongred}{rgb}{0.843137254901961,0.0980392156862745,0.109803921568627}
\definecolor{strongblue}{rgb}{0.172549019607843,0.482352941176471,0.713725490196078}
\definecolor{babyblue}{rgb}{0.54, 0.81, 0.94}
\definecolor{babypink}{rgb}{0.96, 0.76, 0.76}
\definecolor{babygreen}{HTML}{C5E1A5}
\definecolor{stronggreen}{HTML}{26A69A}

\begin{axis}[
axis line style={white!15!black},
legend cell align={left},
legend style={fill opacity=0.8, draw opacity=1, text opacity=1, at={(0.01,0.99)}, anchor=north west, draw=none},
tick align=outside,
tick pos=left,
x grid style={white!80!black},
grid style = {dotted, color = black, line width = 0.5pt},
xlabel={SNR [dB] = Average TX power},
xmajorgrids,
xmin=-12, xmax=32,
xtick style={color=white!15!black},
xtick={-15,-10,-5,0,5,10,15,20,25,30,35},
xticklabels={\(\displaystyle -15\),\(\displaystyle -10\),\(\displaystyle -5\),\(\displaystyle 0\),\(\displaystyle 5\),\(\displaystyle 10\),\(\displaystyle 15\),\(\displaystyle 20\),\(\displaystyle 25\),\(\displaystyle 30\),\(\displaystyle 35\)},
y grid style={white!80!black},
ylabel={Per-user Rate [bpcu]},
ymajorgrids,
ymin=-0.572673111969743, ymax=13.143637256822,
ytick style={color=white!15!black},
ytick={-2,0,2,4,6,8,10,12,14},
yticklabels={\(\displaystyle -2\),\(\displaystyle 0\),\(\displaystyle 2\),\(\displaystyle 4\),\(\displaystyle 6\),\(\displaystyle 8\),\(\displaystyle 10\),\(\displaystyle 12\),\(\displaystyle 14\)},
line width = 0.6pt
]
\addplot [line width=1pt, strongred,  mark=diamond, mark size=1.6, mark options={solid,fill opacity=0}]
table {%
-10 0.0549161062423026
-5 0.141722757908069
0 0.333661394794583
5 0.725973616655588
10 1.45988843415286
15 2.61643372926058
20 4.07042391354971
25 5.6589786321932
30 7.29623423753446
};
\addlegendentry{8 users, $\Delta f=240$~MHz}
\addplot [line width=1pt, strongred, dashed, mark=diamond, mark size=1.6, mark options={solid,fill opacity=0}, forget plot]
table {%
-10 0.0534650104902219
-5 0.138065543405305
0 0.324774122589151
5 0.705561666936932
10 1.4130086924684
15 2.51417971132656
20 3.85115179793781
25 5.18686614160826
30 6.32735842698705
};
\addplot [line width=1pt, strongred, dotted, mark=diamond, mark size=1.6, mark options={solid,fill opacity=0}, forget plot]
table {%
-10 0.0507955411571548
-5 0.130896573612594
0 0.304902659821244
5 0.650075516705057
10 1.26670195017135
15 2.16855524548661
20 3.15955704328696
25 4.01686899725517
30 4.64356119229397
};
\addplot [line width=1pt, stronggreen, mark=star, mark size=1.6, mark options={solid,fill opacity=0}]
table {%
-10 0.145708268997598
-5 0.367168376042691
0 0.845069203920363
5 1.73986281821542
10 3.03675266908759
15 4.5598249279914
20 6.17419829559499
25 7.82009956358612
30 9.4762662895785
};
\addlegendentry{4 users, $\Delta f=240$~MHz}
\addplot [line width=1pt, stronggreen, dashed, mark=star, mark size=1.6, mark options={solid,fill opacity=0}, forget plot]
table {%
-10 0.14324134846612
-5 0.360771916653784
0 0.829374625563676
5 1.7069755236141
10 2.97648825128265
15 4.44409226211697
20 5.93149287025433
25 7.2999921207544
30 8.41892273768506
};
\addplot [line width=1pt, stronggreen, dotted, mark=star, mark size=1.6, mark options={solid,fill opacity=0}, forget plot]
table {%
-10 0.13204462637736
-5 0.333368856965805
0 0.763577319281398
5 1.56151260362527
10 2.70137805611833
15 3.953820447678
20 5.09747343188995
25 6.01216413096652
30 6.66906791082647
};
\addplot [line width=1pt, strongblue, mark=square, mark size=1.6, mark options={solid,fill opacity=0}]
table {%
-10 0.328914780343961
-5 0.812169199211927
0 1.72746560906277
5 3.03791311610544
10 4.56679266636567
15 6.18321364307802
20 7.82978613604606
25 9.48616758529853
30 11.1456792958537
};
\addlegendentry{2 users, $\Delta f=240$~MHz}
\addplot [line width=1pt, strongblue, dashed, mark=square, mark size=1.6, mark options={solid,fill opacity=0}, forget plot]
table {%
-10 0.324182713966161
-5 0.80047020253625
0 1.70399745582665
5 3.00059112114244
10 4.50905141011124
15 6.08239589497182
20 7.62962661104057
25 9.06808437745483
30 10.2991483172127
};
\addplot [line width=1pt, strongblue, dotted, mark=square, mark size=1.6, mark options={solid,fill opacity=0}, forget plot]
table {%
-10 0.298927407602627
-5 0.742863291677549
0 1.59471176748762
5 2.82893386371931
10 4.24802997782102
15 5.66040655798701
20 6.9351703599125
25 7.99531959059729
30 8.81380122792862
};
\addplot [line width=1pt, black]
table {%
-10 0.678703499683111
-5 1.52651539801598
0 2.78755611563926
5 4.29334783864207
10 5.90129952470987
15 7.54505533175647
20 9.20053139819192
25 10.8597553070194
30 12.5201686036951
};
\addlegendentry{1 user, $\Delta f=240$~MHz}
\addplot [line width=1pt, black, dashed, forget plot]
table {%
-10 0.669685935000045
-5 1.51002326555438
0 2.76464447898303
5 4.26673323829566
10 5.87296304939008
15 7.51600837302488
20 9.17122131007599
25 10.8303558002792
30 12.4907400540783
};
\addplot [line width=1pt, black, dotted, forget plot]
table {%
-10 0.620162221520042
-5 1.41795326893615
0 2.63624033076922
5 4.11905824053251
10 5.71777930446425
15 7.35832380399825
20 9.01274900534481
25 10.6716367854913
30 12.3319433970535
};
\end{axis}

\end{tikzpicture}}\\
  \subfloat[][$\Delta f=480$~MHz]{\label{fig: lisa
  c}
\begin{tikzpicture}

\definecolor{strongred}{rgb}{0.843137254901961,0.0980392156862745,0.109803921568627}
\definecolor{strongblue}{rgb}{0.172549019607843,0.482352941176471,0.713725490196078}
\definecolor{babyblue}{rgb}{0.54, 0.81, 0.94}
\definecolor{babypink}{rgb}{0.96, 0.76, 0.76}
\definecolor{babygreen}{HTML}{C5E1A5}
\definecolor{stronggreen}{HTML}{26A69A}

\begin{axis}[
axis line style={white!15!black},
legend cell align={left},
legend style={fill opacity=0.8, draw opacity=1, text opacity=1, at={(0.01,0.99)}, anchor=north west, draw=none},
tick align=outside,
tick pos=left,
x grid style={white!80!black},
grid style = {dotted, color = black, line width = 0.5pt},
xlabel={SNR [dB] = Average TX power},
xmajorgrids,
xmin=-12, xmax=32,
xtick style={color=white!15!black},
xtick={-15,-10,-5,0,5,10,15,20,25,30,35},
xticklabels={\(\displaystyle -15\),\(\displaystyle -10\),\(\displaystyle -5\),\(\displaystyle 0\),\(\displaystyle 5\),\(\displaystyle 10\),\(\displaystyle 15\),\(\displaystyle 20\),\(\displaystyle 25\),\(\displaystyle 30\),\(\displaystyle 35\)},
y grid style={white!80!black},
ylabel={Per-user Rate [bpcu]},
ymajorgrids,
ymin=-0.570911101949923, ymax=13.1026259126346,
ytick style={color=white!15!black},
ytick={-2,0,2,4,6,8,10,12,14},
yticklabels={\(\displaystyle -2\),\(\displaystyle 0\),\(\displaystyle 2\),\(\displaystyle 4\),\(\displaystyle 6\),\(\displaystyle 8\),\(\displaystyle 10\),\(\displaystyle 12\),\(\displaystyle 14\)},
line width = 0.6pt
]
\addplot [line width=1pt, strongred, mark=diamond, mark size=1.6]
table {%
-10 0.0551530043796558
-5 0.143364973131958
0 0.339189257587848
5 0.741055765773624
10 1.49215900451816
15 2.66520366485453
20 4.12819967647115
25 5.72026196486823
30 7.35871013883383
};
\addlegendentry{8 users, $\Delta f=480$~MHz}
\addplot [line width=1pt, strongred, dashed,  mark=diamond, mark size=1.6, mark options={solid,fill opacity=0}, forget plot]
table {%
-10 0.053506450790765
-5 0.139378790319273
0 0.329959189041537
5 0.719897755929332
10 1.44349269655186
15 2.5597180803797
20 3.90328181617028
25 5.23633875182358
30 6.36442425381575
};
\addplot [line width=1pt, strongred, dotted,  mark=diamond, mark size=1.6, mark options={solid,fill opacity=0}, forget plot]
table {%
-10 0.0506133078039181
-5 0.131949983296467
0 0.310621078531513
5 0.666833868093606
10 1.30380474297723
15 2.2295066611149
20 3.2452998735507
25 4.12691465750196
30 4.77708258876491
};
\addplot [line width=1pt, stronggreen, mark=star, mark size=1.6, mark options={solid,fill opacity=0}]
table {%
-10 0.143991680658545
-5 0.365388819355006
0 0.845842332871076
5 1.74498240697029
10 3.04472536818374
15 4.56909565038523
20 6.18393901906586
25 7.8299958366884
30 9.48621247992487
};
\addlegendentry{4 users, $\Delta f=480$~MHz}
\addplot [line width=1pt, stronggreen, dashed, mark=star, mark size=1.6, mark options={solid,fill opacity=0}, forget plot]
table {%
-10 0.141015443910071
-5 0.358170688415626
0 0.829116227070337
5 1.71111857472414
10 2.98356324506507
15 4.45262924984069
20 5.94010258593194
25 7.30570212671591
30 8.41641577549725
};
\addplot [line width=1pt, stronggreen, dotted, mark=star, mark size=1.6, mark options={solid,fill opacity=0}, forget plot]
table {%
-10 0.129869069840375
-5 0.331063776335919
0 0.765926221891112
5 1.57319830908082
10 2.72693930481048
15 3.99935880741332
20 5.16718919294351
25 6.1067784396837
30 6.78673158514316
};
\addplot [line width=1pt, strongblue, mark=square, mark size=1.6, mark options={solid,fill opacity=0}]
table {%
-10 0.322661566970815
-5 0.802028476642517
0 1.71220158135364
5 3.01896624959575
10 4.54616432354509
15 6.16197664851734
20 7.80834778143856
25 9.46466462258082
30 11.1241558075592
};
\addlegendentry{2 users, $\Delta f=480$~MHz}
\addplot [line width=1pt, strongblue, dashed, mark=square, mark size=1.6, mark options={solid,fill opacity=0}, forget plot]
table {%
-10 0.316863877437258
-5 0.788107218121739
0 1.68571217413525
5 2.97824967233923
10 4.48427646259778
15 6.05441835561896
20 7.59493396122679
25 9.01999464237015
30 10.2286614675925
};
\addplot [line width=1pt, strongblue, dotted, mark=square, mark size=1.6, mark options={solid,fill opacity=0}, forget plot]
table {%
-10 0.292528018475527
-5 0.732290095035304
0 1.57783693539432
5 2.80982196116887
10 4.23753455420325
15 5.67098578356905
20 6.97488506650602
25 8.06406159292435
30 8.90472969179717
};
\addplot [line width=1pt, black]
table {%
-10 0.66430663259546
-5 1.50134013253318
0 2.75436064918374
5 4.25635955531473
10 5.8629117576619
15 7.50620110161982
20 9.1615271190491
25 10.820703319114
30 12.4811015028807
};
\addlegendentry{1 user, $\Delta f=480$~MHz}
\addplot [line width=1pt, black, dashed, forget plot]
table {%
-10 0.652978203244253
-5 1.48030219348533
0 2.72503288039503
5 4.22260609770111
10 5.82740020928443
15 7.47008397809303
20 9.12521284438217
25 10.7843261081504
30 12.4447043296512
};
\addplot [line width=1pt, black, dotted, forget plot]
table {%
-10 0.607329258032134
-5 1.39176130517618
0 2.59538815932039
5 4.06763128431579
10 5.66084783404801
15 7.29910812122201
20 8.95271349628156
25 10.6113288982944
30 12.2715478949957
};
\end{axis}

\end{tikzpicture}}
  \caption{Per-user rate results with LISA for perfect (solid), recovered (dashed) and linearly interpolated (dotted) DL CSI knowledge based on fed back DL CSI with compression ratio $ 1/\eta = 40 $.}
\end{figure}

\section{Verifying the Conjecture}
\label{sec: verify_conj}
\noindent The presented results obviously support our fundamental conjecture that learning the channel reconstruction in the UL domain can be ``transfered'' over the frequency gap between UL and DL center frequencies. \vari{In this section, we discuss our intuition behind this conjecture and attempt to falsify the conjecture by examining its limits of applicability.} \\
To investigate the rationale, it is recommended to discuss in more detail the significance of the carrier frequency on the individual parameters of the channel state information. We quickly find that changing the carrier frequency primarily changes the phases of the harmonic signals carried by the corresponding propagation paths. 
\vari{The argumentation is now as follows. For simplicity, we assume the case of a linear uniform array and a single path propagation of a wavefront incident on the antenna array. In this case, the part of the channel vector $ h $ relevant to this discussion can be expressed by $$ h \propto [\alpha^0,\alpha^1,\alpha^2,\dots,\alpha^{N-1}] $$ with $ \alpha = \exp{\left(-j\frac{2\pi df}{c}\sin\theta\right)}$, where $N$ represents the number of antenna elements, $d$ represents the distance between the elements, $f$ represents the carrier frequency of the transmitted narrowband signal, $c$ represents the speed of light, and $\theta$ represents the angle of incidence of the incident wavefront on the antenna array. From this model, one can conclude that small changes $\delta f$ in the carrier frequency can be compensated by small changes $ \delta \theta$ of $\theta$, i.e., the channel vector $h$ is not changed as long as $$ f\sin\theta=(f+\delta f)\sin(\theta+\delta \theta) $$ holds. This means, at least for the special case of the assumed channel model, that from the point of view of the antenna array, for each UL channel vector of a first MT, another second MT can be assumed at a different position with a different angle of arrival of its incident wavefront, but whose DL channel vector is the same as the UL channel vector of user first MT. This is the basis of our intuition: if we observe sufficiently many constellations of channel vectors in the UL and DL domains, we will of course not observe reciprocity between the paired UL and DL vectors of any particular user. However, if we consider the aggregate of a large number of users, that could typically be recorded during standard operation of a BS to generate the training data set that is needed anyway, we will still find that the distribution of the respective UL channels and the distribution of the respective DL channels are nearly equal separately from each other and only when considered by themselves.} \\
\noindent A transfer of the above argumentation for broadband signals and channels on the basis of multipath propagation seems obvious, even if still in the status of a conjecture. This conjecture now states that the totality of all UL channels and the totality of all DL channels have approximately the same properties and can be described with the same or very similar probability distribution, even if individual UL-DL pairs are of course very different. \\ 
The authors are aware of the speculative nature of this reasoning, especially in the case of rich multipath propagation. For this reason, we move to a less qualitative and, on the other hand, more quantitative argumentation. For this purpose, we apply the maximum mean discrepancy measure (MMD) to samples of UL channels on the one hand and DL channels on the other. Along with this, we reformulate our conjecture in that we now assume that samples of UL channels and DL channels of the same scenario represent an identical or similar probability distribution of channel parameters. If this is true, it supports the explanation why a CNN learned with UL channels generalizes on DL channel data.

\subsection{Maximum Mean Discrepancy}
\label{subsec: mmd_def}
\noindent In this section, we introduce the kernel based definition of the so called maximum mean discrepancy (MMD) measure, see~\cite{gretton}, \cite{deepk_paper}, and~\cite{sutherland} for more details. The MMD serves as a means to measure the discrepancy between two probability distributions purely based on its respective samples.

\noindent \textbf{Definition:}
Given a positive definite kernel $k(\cdot,\cdot) = \langle \varphi(\cdot),\varphi(\cdot)\rangle $ 
of
a reproducing kernel Hilbert space (RKHS) $\mathcal{H}_k$ with a feature map $\varphi(\cdot) \in \mathcal{H}_k$, the maximum mean discrepancy (MMD) between two probability distributions $ \mathbb{P}$ and $ \mathbb{Q}$ can be obtained by 
\begin{equation}
\MMD^2(\mathbb{P}, \mathbb{Q},k) \coloneqq {\mathbb{E}[k({p},{p}^\prime) + k({q},{q}^\prime) -2k({p},{q})]}, 
\end{equation}
with random variables $ (p,p^\prime) \sim \mathbb{P} \times \mathbb{P} $ and $ (q,q^\prime) \sim \mathbb{Q} \times \mathbb{Q} $. It follows that $\MMD(\mathbb{P}, \mathbb{Q}, k) = 0$ if and only if $\mathbb{P}= \mathbb{Q}$. If we further assume that we have sample sets $\mathcal{P} \sim \mathbb{P}$ and $\mathcal{Q} \sim \mathbb{Q}$ of equal sample size $n$, an unbiased estimator of the squared MMD for measuring the discrepancy between $ \mathbb{P} $ and $ \mathbb{Q}$ can be obtained by
\begin{equation}
    \widehat{\MMD}^2(\mathcal{P}, \mathcal{Q}, k) \coloneqq \frac{1}{n(n-1)}\sum_{i \neq j}h_{ij},
    \label{eq: mmd}
\end{equation}
where $h_{ij} \coloneqq k(p_i, p_j) + k(q_i, q_j) -k(p_i, q_j) -k(q_i, p_j)$ with $ p_i \in \mathcal{P} $ and $q_i \in \mathcal{Q} $ being the realizations of the random variables $ p \sim \mathbb{P} $ and $ q \sim \mathbb{Q} $.
Following the usual kernel trick, we swap the choice of feature map $ \varphi(\cdot) $ with the decision for a kernel function $ k(\cdot,\cdot) $. The most common choice for a kernel is the Gaussian kernel, i.e.,
$$k(p, q) = \exp\left(-\frac{\norm{p - q}^2}{\sigma_{50}^2}\right),$$ where $p \in \mathcal{P} $
and $q \in \mathcal{Q} $ are two samples drawn from $\mathbb{P}$ and $\mathbb{Q}$ and $\sigma_{50}$ 
corresponds to the 50-percentile (median) distance between elements in the
aggregate sample, as suggested in~\cite{gretton}. 

\begin{figure}[t!]
  \centering
\begin{tikzpicture}

\definecolor{color0}{rgb}{0.54, 0.81, 0.94}

\begin{axis}[
width = 8.4cm,
height = 10cm,
axis line style={white!15!black},
tick align=outside,
tick pos=left,
x grid style={white!80!black},
xmin=0.5, xmax=4.5,
xtick style={color=white!15!black},
xtick={1,2,3,4},
xticklabels={{$120$~MHz}, {$240$~MHz}, {$480$~MHz}, {other cell}},
y grid style={white!80!black},
grid style = {dotted, color = black, line width = 1pt},
ylabel={\(\displaystyle n \cdot \widehat{\textrm{MMD}}^2\)},
ymajorgrids,
ymin=-0.482187881216625, ymax=5.62941535468287,
ytick style={color=white!15!black},
line width = 0.6pt
]
\addplot [black, dotted]
table {%
1 -0.0543286192540293
1 -0.203682562657681
};
\addplot [black, dotted]
table {%
1 0.192329336634572
1 0.454111043715844
};
\addplot [black]
table {%
0.925 -0.203682562657681
1.075 -0.203682562657681
};
\addplot [black]
table {%
0.925 0.454111043715844
1.075 0.454111043715844
};
\addplot [black, dotted]
table {%
2 -0.00670455284890181
2 -0.204387734130285
};
\addplot [black, dotted]
table {%
2 0.14991104507231
2 0.33351242527957
};
\addplot [black]
table {%
1.925 -0.204387734130285
2.075 -0.204387734130285
};
\addplot [black]
table {%
1.925 0.33351242527957
2.075 0.33351242527957
};
\addplot [black, dotted]
table {%
3 0.124807769618138
3 -0.0815205558366561
};
\addplot [black, dotted]
table {%
3 0.280162982269738
3 0.470359496877903
};
\addplot [black]
table {%
2.925 -0.0815205558366561
3.075 -0.0815205558366561
};
\addplot [black]
table {%
2.925 0.470359496877903
3.075 0.470359496877903
};
\addplot [black, dotted]
table {%
4 3.7056374425785
4 2.78487974156466
};
\addplot [black, dotted]
table {%
4 4.39924296584493
4 5.35161520759653
};
\addplot [black]
table {%
3.925 2.78487974156466
4.075 2.78487974156466
};
\addplot [black]
table {%
3.925 5.35161520759653
4.075 5.35161520759653
};
\path [draw=black, fill=color0]
(axis cs:0.9,-0.0543286192540293)
--(axis cs:1.1,-0.0543286192540293)
--(axis cs:1.1,0.192329336634572)
--(axis cs:0.9,0.192329336634572)
--(axis cs:0.9,-0.0543286192540293)
--cycle;
\path [draw=black, fill=color0]
(axis cs:1.9,-0.00670455284890181)
--(axis cs:2.1,-0.00670455284890181)
--(axis cs:2.1,0.14991104507231)
--(axis cs:1.9,0.14991104507231)
--(axis cs:1.9,-0.00670455284890181)
--cycle;
\path [draw=black, fill=color0]
(axis cs:2.9,0.124807769618138)
--(axis cs:3.1,0.124807769618138)
--(axis cs:3.1,0.280162982269738)
--(axis cs:2.9,0.280162982269738)
--(axis cs:2.9,0.124807769618138)
--cycle;
\path [draw=black, fill=color0]
(axis cs:3.9,3.7056374425785)
--(axis cs:4.1,3.7056374425785)
--(axis cs:4.1,4.39924296584493)
--(axis cs:3.9,4.39924296584493)
--(axis cs:3.9,3.7056374425785)
--cycle;
\addplot [semithick, black]
table {%
0.9 0.0412980933885143
1.1 0.0412980933885143
};
\addplot [semithick, black]
table {%
1.9 0.0698144294057623
2.1 0.0698144294057623
};
\addplot [semithick, black]
table {%
2.9 0.185236143488005
3.1 0.185236143488005
};
\addplot [semithick, black]
table {%
3.9 4.01846006617582
4.1 4.01846006617582
};
\end{axis}

\end{tikzpicture}
  \caption{Empirical MMD for UL CSI versus DL CSI.}
  \label{fig: mmd_median_rev_2}
\end{figure}
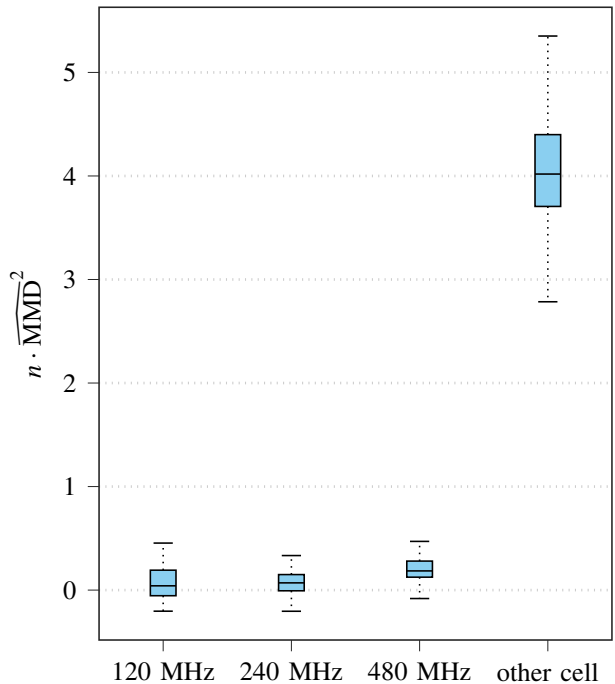

\noindent To apply the MMD to the problem at hand, we compute the discrepancy between the UL CSI sample  $ \mathcal{P}_\text{UL} $ versus the
DL CSI sample $ \mathcal{Q}_\text{DL} $. In particular, we consider four different cases, i.e., the MMD of UL versus DL CSI samples in the same cell for a frequency gap of $\Delta f = 120...480 $ MHz and the MMD in case of a transfer of the UL-learned CNN  to an unknown different environment but for a frequency gap of only $ \Delta f = 120$ MHz. Note that the environment considered here is the same as that we discussed in Section~\ref{sec: cnn_on_another_cell}.

\noindent The corresponding results of the $\widehat{\MMD}^2$ are shown in Figure~\ref{fig: mmd_median_rev_2},
where for each test the $\widehat{\MMD}^2$ has been computed for $100$ paired sets $ \mathcal{P}_\text{UL} $ and $ \mathcal{Q}_\text{DL} $ drawn from the fixed but unknown distributions $ \mathbb{P}_\text{UL} $ and $ \mathbb{Q}_\text{DL} $ of the uplink and downlink channel state information of the respective scenario and cell. The size of each sample is kept constant with $n=1000$ for both $\mathcal{P}_\text{UL}$ and $\mathcal{Q}_\text{DL}$. We can
observe how the MMD for channels that belong to the same cell is much smaller
than the MMD between the UL of the former cell and the DL of a different cell.
Moreover, for the same cell the median value of the box increases as we increase the frequency gap.

\begin{algorithm}[!t]
\caption{True positive rate (TPR) for $H_1$ of the conjecture given a false alarm rate of 5\%}
\begin{algorithmic}
\STATE $ t = 0 $
\FOR{$i =1:\text{\#iterations}$} \STATE $\mathcal{P}_\text{UL} \gets n$ random samples drawn
from $\mathbb{P}_\text{UL} $ \STATE  $\mathcal{Q}_\text{DL} \gets n$ random samples drawn from $\mathbb{Q}_\text{DL} $ \STATE
$d \gets \widehat{\MMD}^2(\mathcal{P}_\text{UL},\mathcal{Q}_\text{DL},k)$
\STATE $\mathcal{D} = \varnothing$ 
\FOR{$j=1:\text{\#permutations}$}
\STATE $\mathcal{P} \gets n$ randomly selected from both $\mathcal{P}_\text{UL} $ and $\mathcal{Q}_\text{DL} $ without replacement \STATE
$\mathcal{Q} \gets n$ randomly selected from both $\mathcal{P}_\text{UL} $ and $\mathcal{Q}_\text{DL} $ without replacement \STATE
$\mathcal{D} \gets \mathcal{D} \cup \widehat{\MMD}^2(\mathcal{P}, \mathcal{Q}, k)$
\ENDFOR \IF{$d > 95$-th percentile of $\mathcal{D}$} \STATE
$t \gets t + 1$ (reject the null hypothesis) \ENDIF \ENDFOR \STATE TPR
$\gets t/\text{\#iterations}$
\end{algorithmic}
\label{algo: perm_test}
\end{algorithm}

\subsection{Hypothesis testing of the conjecture}
\label{subsec: hp_testing}
\noindent A typical feature of the introduced MMD metric is that, being purely data-based, it is per se a random variable and thus its practical application necessarily entails hypothesis testing.
As a consequence, the sole computation of a single MMD metric is not sufficient to prove our conjecture. When having a closer look at the MMD values, we see that they are rather small. Therefore, the question we address in this section is how small the $\widehat{\MMD}^2$ actually needs to be to confirm the hypothesis that the two samples $ \mathcal{P}_\text{UL} $ and $ \mathcal{Q}_{DL} $ of CSI belong to the same distribution. To this end, we follow a method proposed by~\cite{sutherland}, and originally by~\cite{dwass} which is called a permutation test. 
For this purpose, let us define the null hypothesis $H_0$ as the hypothesis that the uplink distribution $\mathbb{P}_\text{UL}$ is equal to the downlink distribution $\mathbb{Q}_\text{DL}$, and let us denote with $H_1$ the alternative hypothesis $\mathbb{P}_\text{UL}\neq \mathbb{Q}_\text{DL}$. If then $H_0$ is true, the samples $ \mathcal{P}_\text{UL} $ and $ \mathcal{Q}_\text{DL} $ are obviously interchangeable. To obtain a meaningful estimation of the conditional distribution of $\widehat{\MMD}^2$ assuming $H_0$ is true, the metric is then repeatedly calculated based on appropriately regenerated data sets $ {\mathcal{P}} $ and $ {\mathcal{Q}} $. These data sets are constructed by uniformly resampling the union of the original data sets $ \mathcal{P}_\text{UL} \cup \mathcal{Q}_\text{DL} $ without replacement, which can also be viewed as splitting the union of the original data sets into two equal sets subsequent to a random permutation. Once a sufficiently large sample of the conditional distribution of the random variable $\widehat{\MMD}^2$ is derived, a false alarm rate for the alternative hypothesis can be introduced and the corresponding decision threshold of the hypothesis test be derived. In contrast, sampling $\widehat{\MMD}^2$ assuming $H_1$ is true cannot be based on resampling or splitting the union of the two data sets under test, but requires multiple original datasets of CSI from the UL and DL domains. 
The main principle of this method is outlined in the Algorithm \ref{algo: perm_test}.

\begin{figure*}[!t]
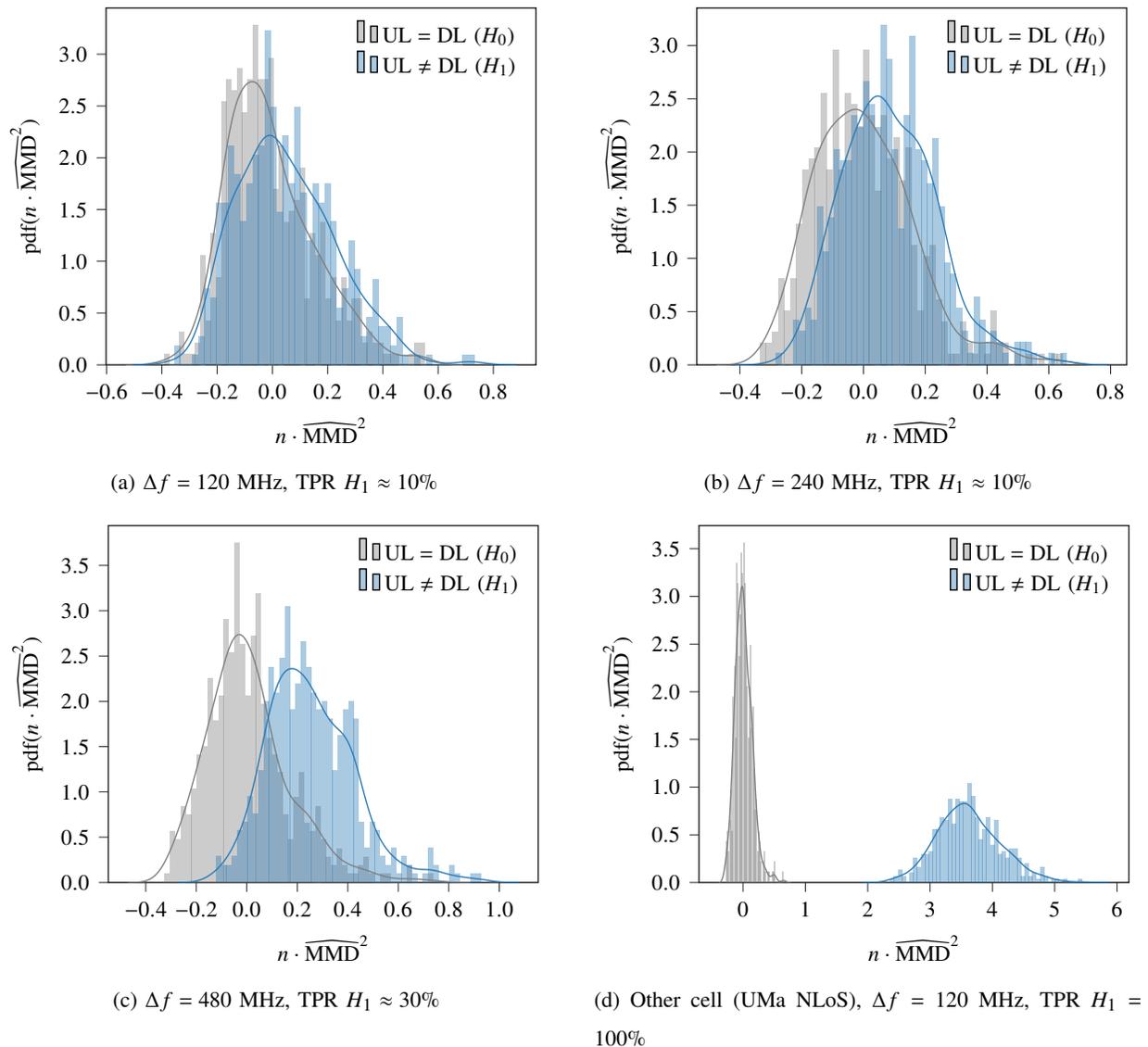

  \centering
  \subfloat[][$\Delta f =120$~MHz, TPR $H_1 \approx
  10\%$]{\label{fig: dist a}\input{distplot_small.tex}}\hspace{0.5cm}
  \subfloat[][$\Delta f =240$~MHz, TPR $H_1 \approx
  10\%$]{\label{fig: dist b}\input{distplot_medium.tex}}\\
  \subfloat[][$\Delta f = 480$~MHz, TPR $H_1 \approx
  30\%$]{\label{fig: dist c}\input{distplot_large.tex}}\hspace{0.5cm}
  \subfloat[][Other cell (UMa NLoS), $\Delta f =120$~MHz, TPR $ H_1=
  100\%$]{\label{fig: dist d}\input{distplot_other_nlos.tex}}
  \caption{Empirical MMD distributions assuming $ H_0 $ or $H_1$ is true, and corresponding TPR for a false alarm rate of $5\%$.}
\end{figure*}

\noindent Finally, evaluating the true positive rate of $H_1$  provides a suitable basis to conclude whether the distributions of CSI at UL and DL frequencies are the same.
In Figures~\ref{fig: dist a}--\ref{fig:
dist d}, we show the empirical distribution of the $\widehat{\MMD}^2$ for both
the two hypothesis $H_0$ and $H_1$. 
In order to define the hypothesis test for a each pair of UL and DL samples $\mathcal{P}_\text{UL} $ and $\mathcal{Q}_\text{DL}$, each of sample size $ n=1000$, we generate $ 500 $ realizations of $\widehat{\MMD}^2$ assuming $ H_0 $ is true (\#permutations) by the respective generation of paired samples $ \mathcal{P} $ and $ \mathcal{Q} $. The determination of the true positive rate (TPR) of the hypothesis test is then based on $100$ test runs (\#iterations) and thus on the corresponding equal number of paired samples $\mathcal{P}_\text{UL} $ and $\mathcal{Q}_\text{DL} $, cf. Algorithm \ref{algo: perm_test}. Each test is designed with respect to a false alarm rate of $5\%$. The resulting TPR is equal with the number of test runs where the MMD indicates a discrepancy between  $\mathcal{P}_\text{UL} $ and $\mathcal{Q}_\text{DL}$ over the total number of tests.

\noindent The obtained results validate our conjecture and confirm that when we have a small gap we can utilize uplink data for training instead of downlink data.
Moreover, in Figure~\ref{fig: dist d} we can clearly see the asymptotic behavior
of the $\widehat{\MMD}^2$ which has been outlined in~\cite{gretton}. Therein, it is reported that the null hypothesis $H_0$ is distributed as an infinite-$\chi^2$-sum, while the alternative hypothesis $H_1$ is normal distributed. For more details, please refer to \cite{gretton}. It should also be mentioned, that the obtained values of $\widehat{\MMD}^2$ depend on the choice of the applied kernel. Consequently, in future work, the analysis of the conjecture could be extended to multiple kernels to eliminate a possible bias of a chosen kernel. On the other hand, due to the universality of the Gaussian kernel, the results obtained are certainly of sufficient quality for the investigations carried out in this work.

\section{Conclusion}
\noindent In this work, we have addressed the problem of channel acquisition for the downlink in FDD communication systems, which typically suffers from the fact that the required CSI cannot be estimated directly at the BS. The classical method for solving this problem is to report back to the base station the channel itself, properties of the channel or quality characteristics derived from it. This work also follows such a general feedback concept by reporting a considerably compressed channel state information to the BS, which subsequently reconstructs the full downlink CSI based on the received feedback. Our novel contribution is that we perform this channel reconstruction using a convolutional neural network and, based on the essential assumption that the CNN can be trained solely on the basis of collected uplink CSI, that the supervisory learning of the CNN is performed without providing downlink data, thus, saving a huge signaling overhead throughout the communication system. On the contrary, it may be assumed that the required training data for purely uplink-based learning is generated anyway during standard uplink operation of the communication link, and thus the provision of the training data does not involve any extra effort. \\
It has been shown that the proposed method clearly outperforms a linear interpolation scheme in terms of conventional performance metrics.  When applying the reconstructed downlink CSI for downlink precoding in a multi-user MIMO communication scenario, it even comes close to the performance achieved when assuming the full knowledge of the downlink CSI at the BS.
The second part of this paper was devoted to strengthening the aforementioned ``transfer learning'' conjecture we have raised by analyzing the equivalence of uplink and downlink CSI for the purpose of learning the weights of the CNN. \\
It is left to future research to further analyze suitable compression schemes of downlink CSI for the intended purpose, the effect of inaccurate channel estimation at the mobile terminal or possible quantization effects. We also intend to address the consideration of the inevitable outdating of the CSI and how it can be accounted for in the learning scheme.

\begin{figure}[!t]
  \centering
\begin{tikzpicture}

\definecolor{color0}{rgb}{0.843137254901961,0.0980392156862745,0.109803921568627}
\definecolor{color1}{rgb}{0.172549019607843,0.482352941176471,0.713725490196078}

\begin{axis}[
axis line style={white!15!black},
legend cell align={left},
legend style={nodes={scale=0.9, transform shape}, fill opacity=0.8, draw opacity=1, text opacity=1, at={(0.01,0.99)}, anchor=north west, draw=none},
tick align=outside,
tick pos=left,
x grid style={white!80!black},
grid style = {dotted, color = black, line width = 0.5pt},
xlabel={SNR [dB] = Average TX power},
xmajorgrids,
xmin=-12, xmax=32,
xtick style={color=white!15!black},
xtick={-15,-10,-5,0,5,10,15,20,25,30,35},
xticklabels={\(\displaystyle -15\),\(\displaystyle -10\),\(\displaystyle -5\),\(\displaystyle 0\),\(\displaystyle 5\),\(\displaystyle 10\),\(\displaystyle 15\),\(\displaystyle 20\),\(\displaystyle 25\),\(\displaystyle 30\),\(\displaystyle 35\)},
y grid style={white!80!black},
ylabel={Per-user Rate [bpcu]},
ymajorgrids,
ymin=-0.320082070462043, ymax=7.56852161218893,
ytick style={color=white!15!black},
ytick={-1,0,1,2,3,4,5,6,7,8},
yticklabels={\(\displaystyle -1\),\(\displaystyle 0\),\(\displaystyle 1\),\(\displaystyle 2\),\(\displaystyle 3\),\(\displaystyle 4\),\(\displaystyle 5\),\(\displaystyle 6\),\(\displaystyle 7\),\(\displaystyle 8\)},
line width = 0.6pt
]
\addplot [line width=1pt, black]
table {%
-10 0.0498816480251544
-5 0.131348004055268
0 0.315298093009574
5 0.694898548267725
10 1.4090482023328
15 2.5462619131671
20 3.98969689587832
25 5.57410648013217
30 7.20994871752298
};
\addlegendentry{perfect channel knowledge}
\addplot [line width=1pt, color0, dashed, mark=star, mark size=1.6, mark options={solid,fill opacity=0}]
table {%
-10 0.0458461293428138
-5 0.121325541102479
0 0.29162464125456
5 0.638849088358753
10 1.27282548460975
15 2.23270606919304
20 3.33363726185574
25 4.31962265932475
30 5.03465421146282
};
\addlegendentry{{CNN, 1.25\% feedback}}
\addplot [line width=1pt, color0, dotted, mark=star, mark size=1.6, mark options={solid,fill opacity=0}]
table {%
-10 0.0384908242039104
-5 0.101095759150664
0 0.23670088804053
5 0.498596184217103
10 0.942929603439154
15 1.54187919016697
20 2.11499222278178
25 2.51425822859674
30 2.73351673636789
};
\addlegendentry{{linear interp., 1.25\% feedback}}
\addplot [line width=1pt, color1, dashed, mark=square, mark size=1.6, mark options={solid,fill opacity=0}]
table {%
-10 0.0473989392446135
-5 0.125130109934885
0 0.300356711788671
5 0.660285042074089
10 1.32680704926979
15 2.36121542395247
20 3.59889459656093
25 4.78916549648987
30 5.74783957878289
};
\addlegendentry{{CNN, 1.875\% feedback}}
\addplot [line width=1pt, color1, dotted, mark=square, mark size=1.6, mark options={solid,fill opacity=0}]
table {%
-10 0.0440236474999325
-5 0.115320093038011
0 0.272911078993576
5 0.580731292398971
10 1.1206859740641
15 1.8887978293017
20 2.69150558762993
25 3.33212015216167
30 3.75300142191464
};
\addlegendentry{linear interp., 1.875\% feedback}

\end{axis}

\end{tikzpicture}
  \caption{Per-user rate results with LISA in a multiuser scenario with 8 users with compression ratios $ 1/\eta \approx 54 $ and $ 1/\eta = 80 $.}
  \label{fig: per_user_diff_compressions}
\end{figure}
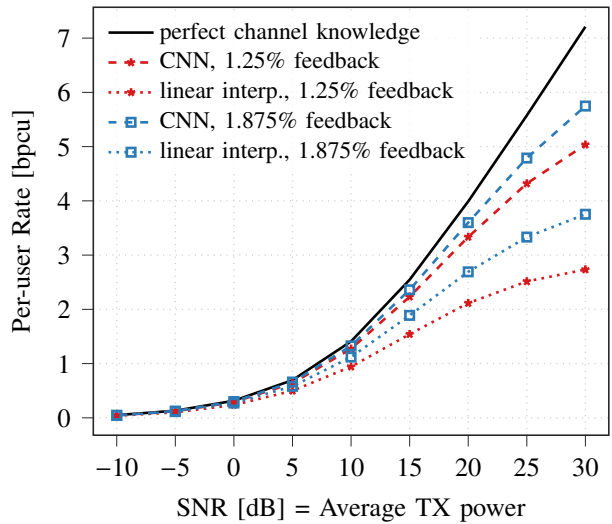

\appendices
\section{Larger compression ratios}
\label{app: larger_cr}
In this section, we investigate the performance degradation when using smaller
quantities of feedback. In particular, we consider the feedback quantities of
$1.25\%$ and $1.875\%$.\\
Once again, the average results of NMSE and cosine similarity for $1.25\%$ and $1.875\%$, which for the sake of brevity we do not report here, have shown that the performance of the DL CSI are
extremely similar to those of the UL test set, although the CNN has been
trained with the sole UL CSI.\\
On the contrary, it might be more interesting for the reader to observe the results in terms of per-user rate in a multi-user scenario with 8 users with the two compression ratios, which are illustrated in Figure~\ref{fig: per_user_diff_compressions}. In particular, we can notice that with both configurations we always improve the linear interpolation scheme.
Note that, in order to slightly improve the results for the $1.25\%$ feedback
case we have changed the values of the dilation of the CNN in Table~\ref{tab:
cnn_lear}, from $[15, 7, 4, 2, 2]$ to $[30, 15, 7, 4, 2]$.



\section{Mask optimization}
\label{app: cae_results}
\begin{figure}[t!]
\vari{
  \centering
  \input{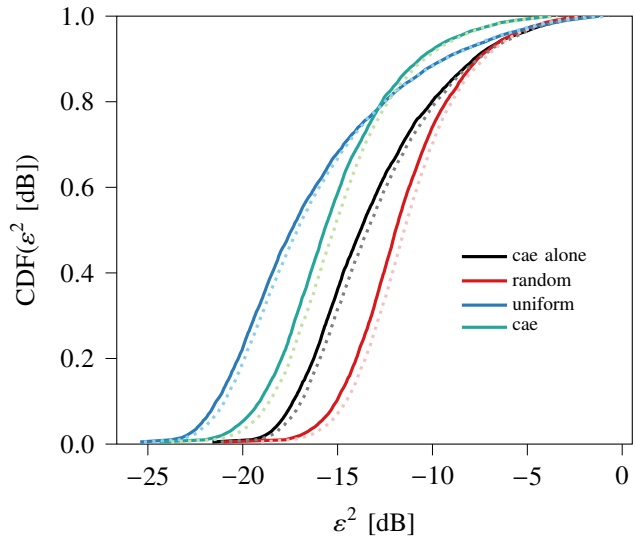}
  \caption{Empirical CDFs of NMSE for fully recovered CSI based on different masks applied to UL test data (solid) and DL test data (dotted).}
  \label{fig: cae_nmse}}
\end{figure}

\noindent \vari{In this Section, we consider the design of the mask $\vM$ which has been introduced in Section~\ref{sec: learn_ul}. To this purpose we decided to utilize the architecture in~\cite{Abid1997} called concrete autoencoder (CAE). The CAE has been applied to a communication problem in~\cite{soltani}, where it has been used to find the most informative locations for pilot transmission in OFDM systems.} \vari{Another very promising study was presented in the aforementioned paper by Mashhadi~\cite{Mashhadi2021b}. However, for the initial investigation of this perspective in our work, we have limited ourselves to performing the optimization of the mask $\vM$ using the method in~\cite{Abid1997}, since it provides an easier integration into our framework. \\ 
In the following, we first summarize the CAE framework, and then we analyze the simulation results.
\subsection{Concrete Autoencoder}
\noindent The CAE is an autoencoder where the first layer called concrete selector layer extracts the $k$ most informative feature of the input $\vx \in \mathbb{R}^d$, where $d \gg k$. Each of the $k$ output neurons is connected to all the input features, through the weights $\vm^{(i)} \in \mathbb{R}^d$, for $i=1, \dots k$:
\begin{equation}
    \vm^{(i)}_j = \frac{\exp{(\log \vect{\alpha}_j + \vg_j)/T}}{\sum_{\ell = 1}^d\exp{(\log \vect{\alpha}_\ell + \vg_\ell)/T}},
\end{equation}
where $\vm^{(i)}_j$ refers to the $j$-th element in of the vector $\vm^{(i)}$, $\vg$ is $d$-dimensional vector sampled from a Gumbel distribution~\cite{Gumbel1954}, $\vect{\alpha} \in \mathbb{R}^d_{> 0}$, and the temperature parameter $T \in (0, \infty)$.
During the training of the CAE, when $T\rightarrow0$ and $\vect{\alpha}$ becomes more sparse, the concrete random variable $\vm^{(i)}$ smoothly approaches the discrete distribution, and outputs a one hot vectors with $\vm^{(i)}_j = 1$ with probability $\vect{\alpha}_j/\sum_{\ell}\vect{\alpha}_\ell$.
\subsection{Simulation results}
\noindent The simulation results obtained with different masks are shown in Figure~\ref{fig: cae_nmse} in terms of NMSE. First of all, one can observe once again that the performances of the UL-trained neural network on DL test data is very close to the performances obtained with UL test data from the same distribution which the recovery network has been trained on. 
The black lines show the performance obtained when the CAE is trained alone without using the proposed CNN, as in~\cite{Abid1997}. The uniform mask together with the CNN which has been discussed in the previous sections outperforms the NMSE that can be obtained by training the CNN with the mask obtained after training the CAE alone.
However, the average value of the NMSE obtained with the CNN trained with respect to the CAE mask ($0.045$ for ``\texttt{cae}'') is better than the average value of the NMSE obtained with the CNN trained with uniform mask ($0.054$ for ``\texttt{uniform}''). This can be expected when having a closer look at Figure~\ref{fig: cae_nmse}: approximately at the value of $-13$~dB the green curves are above the blue ones.
Finally, all these approaches do better than the case in which the CNN is applied to a random mask.
}

\begin{thebibliography}{10}
	\providecommand{\url}[1]{#1}
	\csname url@samestyle\endcsname
	\providecommand{\newblock}{\relax}
	\providecommand{\bibinfo}[2]{#2}
	\providecommand{\BIBentrySTDinterwordspacing}{\spaceskip=0pt\relax}
	\providecommand{\BIBentryALTinterwordstretchfactor}{4}
	\providecommand{\BIBentryALTinterwordspacing}{\spaceskip=\fontdimen2\font plus
		\BIBentryALTinterwordstretchfactor\fontdimen3\font minus
		\fontdimen4\font\relax}
	\providecommand{\BIBforeignlanguage}[2]{{%
			\expandafter\ifx\csname l@#1\endcsname\relax
			\typeout{** WARNING: IEEEtran.bst: No hyphenation pattern has been}%
			\typeout{** loaded for the language `#1'. Using the pattern for}%
			\typeout{** the default language instead.}%
			\else
			\language=\csname l@#1\endcsname
			\fi
			#2}}
	\providecommand{\BIBdecl}{\relax}
	\BIBdecl
	
	\bibitem{marzetta}
	T.~L. {Marzetta}, ``{Noncooperative Cellular Wireless with Unlimited Numbers of
		Base Station Antennas},'' \emph{{IEEE Transactions on Wireless
			Communications}}, vol.~{9}, no.~{11}, pp. {3590--3600}, {2010}.
	
	\bibitem{sanguinetti2019massive}
	L.~{Sanguinetti}, E.~{Bj{\"o}rnson}, and J.~{Hoydis}, ``{Toward Massive MIMO
		2.0: Understanding Spatial Correlation, Interference Suppression, and Pilot
		Contamination},'' \emph{IEEE Transactions on Communications}, vol.~68, no.~1,
	pp. 232--257, 2020.
	
	\bibitem{2019massive}
	E.~Bj{\"o}rnson, L.~Sanguinetti, H.~Wymeersch, J.~Hoydis, and T.~L. Marzetta,
	``Massive {MIMO} is a reality--{W}hat is next? five promising research
	directions for antenna arrays,'' \emph{Digital Signal Processing}, vol.~94,
	pp. 3 -- 20, 2019, special Issue on Source Localization in Massive MIMO.
	
	\bibitem{8542957}
	M.~{Barzegar Khalilsarai}, S.~{Haghighatshoar}, X.~{Yi}, and G.~{Caire},
	``{FDD} massive {MIMO} via {UL/DL} channel covariance extrapolation and
	active channel sparsification,'' \emph{IEEE Transactions on Wireless
		Communications}, vol.~18, no.~1, pp. 121--135, 2019.
	
	\bibitem{UtNo99}
	W.~Utschick and J.~A. Nossek, ``Downlink beamforming for {FDD} mobile radio
	systems based on spatial covariances,'' in \emph{Proceedings of the European
		Wireless 99 \& ITG Mobile Communications}, Munich, Germany, 1999, pp. 65--67.
	
	\bibitem{ArDoCaYaHoBr19}
	M.~Arnold, S.~D{\"{o}}rner, S.~Cammerer, S.~Yan, J.~Hoydis, and S.~ten Brink,
	``Enabling {FDD} massive {MIMO} through deep learning-based channel
	prediction,'' \emph{CoRR}, vol. abs/1901.03664, 2019.
	
	\bibitem{alk2019deep}
	M.~{Alrabeiah} and A.~{Alkhateeb}, ``{Deep Learning for TDD and {FDD} Massive
		MIMO: Mapping Channels in Space and Frequency},'' in \emph{2019 53rd Asilomar
		Conference on Signals, Systems, and Computers}, 2019, pp. 1465--1470.
	
	\bibitem{8764345}
	J.~{Wang}, Y.~{Ding}, S.~{Bian}, Y.~{Peng}, M.~{Liu}, and G.~{Gui}, ``{UL-CSI}
	data driven deep learning for predicting {DL-CSI} in cellular {FDD}
	systems,'' \emph{IEEE Access}, vol.~7, pp. 96\,105--96\,112, 2019.
	
	\bibitem{han2020deep}
	Y.~{Han}, M.~{Li}, S.~{Jin}, C.~K. {Wen}, and X.~{Ma}, ``{Deep Learning-Based
		{FDD} Non-Stationary Massive MIMO Downlink Channel Reconstruction},''
	\emph{IEEE Journal on Selected Areas in Communications}, vol.~38, no.~9, pp.
	1980--1993, 2020.
	
	\bibitem{safari}
	M.~S. {Safari}, V.~{Pourahmadi}, and S.~{Sodagari}, ``{Deep UL2DL: Data-Driven
		Channel Knowledge Transfer From Uplink to Downlink},'' \emph{IEEE Open
		Journal of Vehicular Technology}, vol.~1, pp. 29--44, 2020.
	
	\bibitem{me}
	V.~{Rizzello}, I.~{Brayek}, M.~{Joham}, and W.~{Utschick}, ``{Learning the
		Channel State Information Across the Frequency Division Gap in Wireless
		Communications},'' in \emph{WSA 2020; 24th International ITG Workshop on
		Smart Antennas}, 2020, pp. 1--6.
	
	\bibitem{4641946}
	D.~J. {Love}, R.~W. {Heath}, V.~K. {N. Lau}, D.~{Gesbert}, B.~D. {Rao}, and
	M.~{Andrews}, ``An overview of limited feedback in wireless communication
	systems,'' \emph{IEEE Journal on Selected Areas in Communications}, vol.~26,
	no.~8, pp. 1341--1365, 2008.
	
	\bibitem{8322184}
	C.~{Wen}, W.~{Shih}, and S.~{Jin}, ``Deep learning for massive {MIMO CSI}
	feedback,'' \emph{IEEE Wireless Communications Letters}, vol.~7, no.~5, pp.
	748--751, 2018.
	
	\bibitem{8638509}
	Z.~{Liu}, L.~{Zhang}, and Z.~{Ding}, ``Exploiting bi-directional channel
	reciprocity in deep learning for low rate massive {MIMO CSI} feedback,''
	\emph{IEEE Wireless Communications Letters}, vol.~8, no.~3, pp. 889--892,
	2019.
	
	\bibitem{9090892}
	------, ``An efficient deep learning framework for low rate massive {MIMO CSI}
	reporting,'' \emph{IEEE Transactions on Communications}, vol.~68, no.~8, pp.
	4761--4772, 2020.
	
	\bibitem{8972904}
	J.~{Guo}, C.~{Wen}, S.~{Jin}, and G.~Y. {Li}, ``Convolutional neural
	network-based multiple-rate compressive sensing for massive {MIMO CSI}
	feedback: Design, simulation, and analysis,'' \emph{IEEE Transactions on
		Wireless Communications}, vol.~19, no.~4, pp. 2827--2840, 2020.
	
	\bibitem{9279228}
	J.~{Guo}, C.~K. {Wen}, and S.~{Jin}, ``Deep learning-based {CSI} feedback for
	beamforming in single- and multi-cell massive {MIMO} systems,'' \emph{IEEE
		Journal on Selected Areas in Communications}, pp. 1--1, 2020.
	
	\bibitem{9347820}
	F.~{Sohrabi}, K.~M. {Attiah}, and W.~{Yu}, ``Deep learning for distributed
	channel feedback and multiuser precoding in {FDD} massive {MIMO},''
	\emph{IEEE Transactions on Wireless Communications}, pp. 1--1, 2021.
	
	\bibitem{Liang2020}
	P.~Liang, J.~Fan, W.~Shen, Z.~Qin, and G.~Y. Li, ``{Deep learning and
		compressive sensing-based csi feedback in fdd massive mimo systems},''
	\emph{IEEE Transactions on Vehicular Technology}, vol.~69, no.~8, pp.
	9217--9222, 2020.
	
	\bibitem{Mashhadi2021b}
	M.~B. Mashhadi and D.~Gunduz, ``{Pruning the Pilots: Deep Learning-Based Pilot
		Design and Channel Estimation for MIMO-OFDM Systems},'' \emph{IEEE
		Transactions on Wireless Communications}, vol. 1276, no.~c, pp. 1--12, 2021.
	
	\bibitem{quad}
	S.~{Jaeckel}, L.~{Raschkowski}, F.~{Burkhardt}, and L.~{Thiele}, ``{Efficient
		Sum-of-Sinusoids-Based Spatial Consistency for the 3GPP New-Radio Channel
		Model},'' in \emph{2018 IEEE Globecom Workshops (GC Wkshps)}, 2018, pp. 1--7.
	
	\bibitem{quad2}
	M.~{Kurras}, S.~{Dai}, S.~{Jaeckel}, and L.~{Thiele}, ``{Evaluation of the
		Spatial Consistency Feature in the 3GPP Geometry-Based Stochastic Channel
		Model},'' in \emph{2019 IEEE Wireless Communications and Networking
		Conference (WCNC)}, 2019, pp. 1--6.
	
	\bibitem{Mashhadi2021a}
	M.~B. Mashhadi, Q.~Yang, and D.~Gunduz, ``{Distributed Deep Convolutional
		Compression for Massive MIMO CSI Feedback},'' \emph{IEEE Transactions on
		Wireless Communications}, vol.~20, no.~4, pp. 2621--2633, 2021.
	
	\bibitem{YuKo15}
	F.~Yu and V.~Koltun, ``Multi-scale context aggregation by dilated
	convolutions,'' in \emph{4th International Conference on Learning
		Representations, {ICLR} 2016, San Juan, Puerto Rico, May 2-4, 2016,
		Conference Track Proceedings}, Y.~Bengio and Y.~LeCun, Eds., 2016.
	
	\bibitem{tensorflow2015}
	``{TensorFlow}: Large-scale machine learning on heterogeneous systems,'' 2015,
	software available from tensorflow.org.
	
	\bibitem{adam}
	D.~P. Kingma and J.~Ba, ``{Adam: {A} Method for Stochastic Optimization},'' in
	\emph{3rd International Conference on Learning Representations, {ICLR} 2015,
		San Diego, CA, USA, May 7-9, 2015, Conference Track Proceedings}, Y.~Bengio
	and Y.~LeCun, Eds., 2015.
	
	\bibitem{1033876}
	S.~{Coleri}, M.~{Ergen}, A.~{Puri}, and A.~{Bahai}, ``Channel estimation
	techniques based on pilot arrangement in {OFDM} systems,'' \emph{IEEE
		Transactions on Broadcasting}, vol.~48, no.~3, pp. 223--229, 2002.
	
	\bibitem{GuUtDi09b}
	C.~Guthy, W.~Utschick, and G.~Dietl, ``Low complexity linear zero-forcing for
	the {MIMO} broadcast channel,'' \emph{{IEEE Journal on Selected Topics in
			Signal Processing}}, vol.~3, no.~6, pp. 1106--1117, December 2009.
	
	\bibitem{lisa}
	W.~{Utschick}, C.~{St{\"o}ckle}, M.~{Joham}, and J.~{Luo}, ``{Hybrid LISA
		Precoding for Multiuser Millimeter-Wave Communications},'' \emph{IEEE
		Transactions on Wireless Communications}, vol.~17, no.~2, pp. 752--765, 2018.
	
	\bibitem{1705034}
	P.~{Tejera}, W.~{Utschick}, G.~{Bauch}, and J.~A. {Nossek}, ``{Subchannel
		Allocation in Multiuser Multiple-Input-Multiple-Output Systems},'' \emph{IEEE
		Transactions on Information Theory}, vol.~52, no.~10, pp. 4721--4733, 2006.
	
	\bibitem{Mashhadi2021}
	M.~B. Mashhadi, M.~Jankowski, T.~Tung, S.~Kobus, and D.~G{\"{u}}nd{\"{u}}z,
	``Federated mmwave beam selection utilizing {LIDAR} data,'' \emph{CoRR}, vol.
	abs/2102.02802, 2021.
	
	\bibitem{gretton}
	A.~Gretton, K.~M. Borgwardt, M.~J. Rasch, B.~Sch{\"{o}}lkopf, and A.~J. Smola,
	``{A Kernel Two-Sample Test},'' \emph{J. Mach. Learn. Res.}, vol.~13, pp.
	723--773, 2012.
	
	\bibitem{deepk_paper}
	F.~Liu, W.~Xu, J.~Lu, G.~Zhang, A.~Gretton, and D.~J. Sutherland, ``{Learning
		Deep Kernels for Non-Parametric Two-Sample Tests},'' in \emph{Proceedings of
		the 37th International Conference on Machine Learning, {ICML} 2020, 13-18
		July 2020, Virtual Event}, ser. Proceedings of Machine Learning Research,
	vol. 119.\hskip 1em plus 0.5em minus 0.4em\relax {PMLR}, 2020, pp.
	6316--6326.
	
	\bibitem{sutherland}
	D.~J. Sutherland, H.~Tung, H.~Strathmann, S.~De, A.~Ramdas, A.~J. Smola, and
	A.~Gretton, ``{Generative Models and Model Criticism via Optimized Maximum
		Mean Discrepancy},'' in \emph{5th International Conference on Learning
		Representations, {ICLR} 2017, Toulon, France, April 24-26, 2017, Conference
		Track Proceedings}.\hskip 1em plus 0.5em minus 0.4em\relax OpenReview.net,
	2017.
	
	\bibitem{dwass}
	M.~Dwass, ``Modified randomization tests for nonparametric hypotheses,''
	\emph{The Annals of Mathematical Statistics}, vol.~28, no.~1, pp. 181--187,
	1957.
	
	\bibitem{Abid1997}
	M.~F. Balin, A.~Abid, and J.~Y. Zou, ``Concrete autoencoders: Differentiable
	feature selection and reconstruction,'' in \emph{Proceedings of the 36th
		International Conference on Machine Learning, {ICML} 2019, 9-15 June 2019,
		Long Beach, California, {USA}}, ser. Proceedings of Machine Learning
	Research, K.~Chaudhuri and R.~Salakhutdinov, Eds., vol.~97.\hskip 1em plus
	0.5em minus 0.4em\relax {PMLR}, 2019, pp. 444--453.
	
	\bibitem{soltani}
	M.~Soltani, V.~Pourahmadi, and H.~Sheikhzadeh, ``Pilot pattern design for deep
	learning-based channel estimation in {OFDM} systems,'' \emph{{IEEE} Wirel.
		Commun. Lett.}, vol.~9, no.~12, pp. 2173--2176, 2020.
	
	\bibitem{Gumbel1954}
	E.~J. Gumbel, \emph{\BIBforeignlanguage{eng}{Statistical theory of extreme
			values and some practical applications; a series of lectures}}, ser. Applied
	mathematics series ; 33.\hskip 1em plus 0.5em minus 0.4em\relax Washington:
	U.S. Govt. Print. Office, 1954.
	
\end{thebibliography}
\end{document}